\documentclass[english]{IEEEtran}
\usepackage[T1]{fontenc}
\usepackage[latin9]{inputenc}
\usepackage{color}
\usepackage{babel}
\usepackage{array}
\usepackage{float}
\usepackage{multirow}
\usepackage{amsmath}
\usepackage{amsthm}
\usepackage{amssymb}
\usepackage{graphicx}
\usepackage[unicode=true]
 {hyperref}

\makeatletter

\providecommand{\tabularnewline}{\\}
\floatstyle{ruled}
\newfloat{algorithm}{tbp}{loa}
\providecommand{\algorithmname}{Algorithm}
\floatname{algorithm}{\protect\algorithmname}

\theoremstyle{plain}
\newtheorem{thm}{\protect\theoremname}
\theoremstyle{remark}
\newtheorem{rem}[thm]{\protect\remarkname}
\theoremstyle{plain}
\newtheorem{lem}[thm]{\protect\lemmaname}

\usepackage{algorithmic}
\usepackage{caption}
\usepackage{cite}

\makeatother

\providecommand{\lemmaname}{Lemma}
\providecommand{\remarkname}{Remark}
\providecommand{\theoremname}{Theorem}

\begin{document}

\title{Efficient and Scalable Parametric High-Order Portfolios Design via
the Skew-$t$ Distribution}

\author{Xiwen~Wang, Rui~Zhou, Jiaxi~Ying, and Daniel~P.~Palomar,~\IEEEmembership{Fellow,~IEEE}\thanks{This work was supported by the Hong Kong GRF 16207820 research grant,
the National Nature Science Foundation of China (NSFC) under Grant
62201362, and the Shenzhen Science and Technology Program (Grant No.
RCBS20221008093126071). (Corresponding author: Rui Zhou and Jiaxi
Ying.)}\thanks{Xiwen~Wang is with the Department of Electronic and Computer Engineering,
Hong Kong University of Science and Technology, Clear Water Bay, Kowloon,
Hong Kong (e-mail: \protect\href{mailto:xwangew@connect.ust.hk}{xwangew@connect.ust.hk}). }\thanks{Rui Zhou is with the Shenzhen Research Institute of Big Data, Shenzhen,
China (email: \protect\href{mailto:rui.zhou@sribd.cn}{rui.zhou@sribd.cn})}\thanks{Jiaxi Ying is with the Department of Mathematics, Hong Kong University
of Science and Technology, Clear Water Bay, Kowloon, Hong Kong (e-mail:
\protect\href{mailto:jx.ying@connect.ust.hk}{jx.ying@connect.ust.hk}).}\thanks{Daniel P. Palomar is with the Department of Electronic and Computer
Engineering and Department of Industrial Engineering and Decision
Analytics, Hong Kong University of Science and Technology, Kowloon,
Hong Kong (e-mail: \protect\href{mailto:palomar@ust.hk}{palomar@ust.hk}).}}
\maketitle
\begin{abstract}
Since Markowitz\textquoteright s mean-variance framework, optimizing
a portfolio that strikes a trade-off between maximizing profit and
minimizing risk has been ubiquitous in the financial industry. Initially,
profit and risk were measured by the first two moments of the portfolio's
return, a.k.a. the mean and variance, which are sufficient to characterize
a Gaussian distribution. However, it is broadly believed that the
first two moments are not enough to capture the characteristics of
the returns' behavior, which have been recognized to be asymmetric
and heavy-tailed. Although portfolio designs involving the third and
fourth moments, i.e., skewness and kurtosis, have been demonstrated
to outperform the conventional mean-variance framework, they present
non-trivial challenges. Specifically, in the classical framework,
the memory and computational cost of computing the skewness and kurtosis
grow sharply with the number of assets. To alleviate the difficulty
in high-dimensional problems, we consider an alternative expression
for high-order moments based on parametric representations via a generalized
hyperbolic skew-$t$ distribution. Then, we reformulate the high-order
portfolio optimization problem as a fixed-point problem and propose
a robust fixed-point acceleration algorithm that solves the problem
in an efficient and scalable manner. Empirical experiments also attest
to the efficiency of our proposed high-order portfolio optimization
framework, which presents low complexity and significantly outperforms
the state-of-the-art methods by $2\sim4$ orders of magnitude. 
\end{abstract}


\section{\textcolor{black}{Introduction}}

Modern portfolio theory (MPT), pioneered by Harry Markowitz \cite{markowitz1952portfolio},
strives to reaching a trade-off between minimizing the risk of the
portfolio and maximizing its profit. For the convenience of modeling
the profit and risk, the assets' returns are conventionally assumed
to follow a Gaussian distribution. The Gaussian distribution was embraced
in early research for a number of reasons \cite{markowitz1991foundations}.
First of all, it is straightforward to describe the data using the
Gaussian distribution. The mean vector $\boldsymbol{\mu}$ and covariance
matrix $\boldsymbol{\Sigma}$, which are the parameters of the Gaussian
distribution, can be obtained via numerous estimation methods. Moreover,
the mathematical expression of profit and risk are henceforth simple
enough such that the resultant portfolio designs are convenient from
the perspective of optimization. However, the mean and variance, a.k.a.
the first- and second-order moments, are usually not sufficient to
capture the characteristics of the assets' returns \cite{adcock2015skewed,resnick2007heavy}.
It is widely acknowledged that empirical observations of stock data
exhibit asymmetry and fat tails that can be barely described by a
Gaussian distribution \cite{kolm201460,de2021graphical,jondeau2003conditional,de2022learning}.
In light of these deficiencies, a number of empirical evidence advocates
the incorporation of the high-order moments into portfolio design
\cite{bradley2003financial,rosenberg2006general}. 

The concerns of skewness and kurtosis, a.k.a. third- and fourth-order
moments, have been raised for decades \cite{gaurav2013effect}. Typically,
higher skewness is preferred as it reduces extreme values on the side
of losses and increases them on the side of gains. Whereas the kurtosis
measures dispersion which is something undesirable that increases
the uncertainty of returns \cite{maringer2009global,scott1980direction,decarlo1997meaning}.
A detailed discussion can be found in \cite{scott1980direction}.
Therefore, portfolio designs should also aspire to achieve high skewness
and low kurtosis. This trade-off was then naturally formulated as
a mean-variance-skewness-kurtosis (MVSK) framework \cite{jondeau2006optimal}.

Although there are many compelling advantages of involving skewness
and kurtosis \cite{harvey2010portfolio,he2014multi}, solving high-order
portfolio optimization problems is non-trivial. Given a problem formulation
to specify the trade-off, a typical high-order portfolio design consists
of a model to characterize the high-order moments and optimization
algorithms to solve the problem. Each of these modules can be a limiting
factor in the overall practicability of the framework. In this paper,
we start from the classical MVSK problem formulation. Then, the first
fundamental problem is how to model the skewness and kurtosis of the
portfolio return. The conventional approach models the skewness and
kurtosis via the vanilla co-skewness matrix $\boldsymbol{\Phi}\in\mathbb{R}^{N\times N^{2}}$
and co-kurtosis matrix $\boldsymbol{\Psi}\in\mathbb{R}^{N\times N^{3}}$.
However, this non-parametric modeling suffers a lot from the dimensionality
problem \cite{martellini2010improved}, which might not be critical
on variance but is severely exacerbated on estimating skewness and
kurtosis. For example, to obtain $\boldsymbol{\Phi}$ and $\boldsymbol{\Psi}$
when $N=100$, we need to estimate more than $170$ thousand and $4$
million parameters, respectively. As the number of parameters is significantly
larger than the number of samples, the estimation error is inevitably
large \cite{jondeau2016asymmetry}. In addition, the storage burden
is also exceptionally heavy. Any mathematical manipulations involving
$\boldsymbol{\Phi}$ and $\boldsymbol{\Psi}$ would demand prohibitive
computational resources and are thus not applicable to high-dimensional
problems. 

Apart from the high computational cost due to the matrices $\boldsymbol{\Phi}$
and $\boldsymbol{\Psi}$, the third moment of the portfolio return
is non-convex, making it difficult to optimize. The existing methods
in the literature can be roughly classified into three major categories:
zeroth-order, first-order, and second-order methods. The zeroth-order
methods \cite{liu2020primer} usually iteratively improve the objective
values via repetitive function evaluations. For instance, the differential
evolution \cite{babu2003differential} and genetic algorithms \cite{kshatriya2018genetic}
iteratively improve solutions via searching in the feasible region.
Usually, zeroth-order searching algorithms are often criticized for
their mediocre performance on the computational cost. The first-order
methods make use of the first-order derivative of the objective function.
Some classical examples include the difference-of-convex (DC) algorithms
\cite{dinh2011efficient,niu2019higher} and some Majorization-Minimization
algorithms \cite{zhou2021solving}. However, the first-order methods
may need quite a large number of iterations to converge. In contrast,
the second-order methods improve the description of the descent direction
by incorporating the second-order derivative of the objective function.
For example, the Q-MVSK algorithm \cite{zhou2021solving} presents
a significantly faster convergence rate than the first-order methods.
However, the per-iteration cost of second-order methods is prohibitive
as computing the Hessian has dramatically high complexity. 

In summary, due to the computationally expensive modeling of high-order
moments and the absence of practical optimization algorithms, the
current MVSK framework can only produce high-order portfolios in low-dimensional
problems. To address these limitations, in this paper, we present
a novel high-order portfolio design framework that is both efficient
and scalable. Our contributions are mainly twofold:
\begin{enumerate}
\item We adopt a parametric model to significantly reduce the memory and
computational cost of obtaining the high-order moments of the portfolio
return. The proposed method accommodates the high-dimensional scenarios
by fitting the data via a generalized hyperbolic skew-$t$ distribution. 
\item We propose a practical algorithm based on a robust fixed point acceleration
strategy to solve the high-order portfolios. The numerical experiments
demonstrate that the proposed algorithms are significantly more efficient
and scalable than the state-of-the-art solvers. 
\end{enumerate}
The structure of this paper is as follows. In Section \ref{sec:Problem-formulation},
we first introduce the high-order portfolio optimization problems
and illustrate the current difficulties. In Section \ref{sec:Modeling-higher-order-using-MVST},
we present an efficient approach to compute the skewness and kurtosis
using a generalized hyperbolic skew-$t$ distribution. The parametric
distribution allows a faster way of computing high-order moments of
the portfolio. In Section \ref{sec:Solving-MVSK-Portfolios}, we propose
efficient algorithms to solve the MVSK portfolios based on fixed point
acceleration strategy. Additionally, in Section \ref{sec:Solving-MVSK-Tilting-Portfolios},
we show that the proposed algorithm can be easily generalized into
the MVSK-Tilting portfolio. Then, we elaborate the performance of
proposed high-order portfolio design framework in Section \ref{sec:Numerical-Simulations}.
Finally, we summarize the conclusions in section \ref{sec:Conclusion}.

\section{Problem formulations \label{sec:Problem-formulation} }

\subsection{MVSK Portfolios}

Let $\mathbf{r}\in\mathbb{R}^{N}$ denote the log-returns of $N$
assets and $\mathbf{w}\in\mathbb{R}^{N}$ denote the portfolio weights.
The classical mean-variance portfolio optimization problem is formulated
as 

\begin{equation}
\begin{array}{ll}
\underset{\mathbf{w}}{\mathsf{minimize}}\, & -\phi_{1}\left(\mathbf{w}\right)+\lambda\phi_{2}\left(\mathbf{w}\right)\\
\mathsf{subject}\text{ }\mathsf{to} & \mathbf{w}\in\mathcal{W},
\end{array}\label{eq:MVO}
\end{equation}
where $\phi_{1}\left(\mathbf{w}\right)$ refers to the first central
moment, a.k.a. the mean of the portfolio return, i.e., 
\begin{equation}
\phi_{1}\left(\mathbf{w}\right)=\mathbb{E}\left[\mathbf{w}^{T}\mathbf{r}\right],\label{eq:phi1w}
\end{equation}
$\phi_{2}\left(\mathbf{w}\right)$ is the second central moment, which
is the variance of the portfolio return, i.e., 
\begin{equation}
\phi_{2}\left(\mathbf{w}\right)=\mathbf{\mathbb{E}}\left[\left(\mathbf{w}^{T}\mathbf{r}-\mathbb{E}\left[\mathbf{w}^{T}\mathbf{r}\right]\right)^{2}\right],\label{eq:phi2w}
\end{equation}
$\lambda>0$ is a risk-aversion coefficient, and $\mathcal{W}$ represents
the feasible set of the portfolio weights. In the paper, we consider
no-shorting. Therefore, $\mathcal{W}$ is a unit simplex denoted as
\begin{equation}
\mathcal{W}=\left\{ \mathbf{w}\left|\mathbf{1}^{T}\mathbf{w}=1,\mathbf{w}\geq\mathbf{0}\right.\right\} .\label{eq:definition_of_w}
\end{equation}

Now, we incorporate the third and fourth central moments of the portfolio
return, i.e.,
\begin{equation}
\begin{aligned}\phi_{3}\left(\mathbf{w}\right) & =\mathbf{\mathbb{E}}\left[\left(\mathbf{w}^{T}\mathbf{r}-\mathbb{E}\left[\mathbf{w}^{T}\mathbf{r}\right]\right)^{3}\right],\\
\phi_{4}\left(\mathbf{w}\right) & =\mathbf{\mathbb{E}}\left[\left(\mathbf{w}^{T}\mathbf{r}-\mathbb{E}\left[\mathbf{w}^{T}\mathbf{r}\right]\right)^{4}\right],
\end{aligned}
\label{eq:phi34w}
\end{equation}
into the portfolio selection. This directly extends the mean-variance
portfolio into a mean-variance-skewness-kurtosis (MVSK) portfolio,
formulated as follows
\begin{equation}
\begin{array}{ll}
\underset{\mathbf{w}}{\mathsf{minimize}} & f\left(\mathbf{w}\right)=-\lambda_{1}\phi_{1}\left(\mathbf{w}\right)+\lambda_{2}\phi_{2}\left(\mathbf{w}\right)\\
 & \quad\quad\quad\,\,\,\,-\lambda_{3}\phi_{3}\left(\mathbf{w}\right)+\lambda_{4}\phi_{4}\left(\mathbf{w}\right)\\
\mathsf{subject}\text{ }\mathsf{to} & \mathbf{w}\in\mathcal{W},
\end{array}\label{eq:MVSK}
\end{equation}
where $\lambda_{1},\lambda_{2},\lambda_{3},\lambda_{4}$ are the non-negative
parameters controlling the relative importance of individual moments. 

\subsection{Current Difficulties \label{subsec:Current-difficulties}}

Among many difficulties regarding high-order portfolio designs, the
most fundamental bottleneck is the prohibitive cost of computing high-order
central moments using the non-parametric representation. Namely, the
conventional way applies the following formulas to characterize the
co-skewness and co-kurtosis matrices, 
\begin{align}
\boldsymbol{\Phi} & =\mathbb{E}\left[\left(\mathbf{r}-\boldsymbol{\mu}\right)\left(\mathbf{r}-\boldsymbol{\mu}\right)\otimes\left(\mathbf{r}-\boldsymbol{\mu}\right)\right],\nonumber \\
\boldsymbol{\Psi} & =\mathbb{E}\left[\left(\mathbf{r}-\boldsymbol{\mu}\right)\left(\mathbf{r}-\boldsymbol{\mu}\right)\otimes\left(\mathbf{r}-\boldsymbol{\mu}\right)\otimes\left(\mathbf{r}-\boldsymbol{\mu}\right)\right],\label{eq:phianspsi}
\end{align}
where $\boldsymbol{\mu}=\mathbb{E}\left[\mathbf{r}\right]$. As shown
in Table \ref{tab:Conventional-non-parametric-appr}, the costs for
storing $\boldsymbol{\Phi}$ and $\boldsymbol{\Psi}$ have a high
complexity. This means that we may not be able to set up these matrices
when the problem dimension is large.
\begin{table}
\caption{Conventional non-parametric representations of high-order moments.
\label{tab:Conventional-non-parametric-appr}}

\centering{}%
\begin{tabular}{|c|c|c|}
\hline 
\multirow{2}{*}{} & Number of parameters  & Memory \tabularnewline
 & to estimate & complexity\tabularnewline
\hline 
Co-skewness $\boldsymbol{\Phi}$ & $\frac{1}{6}N\left(N+1\right)\left(N+2\right)$ & $\mathcal{O}\left(N^{3}\right)$\tabularnewline
\hline 
Co-kurtosis $\boldsymbol{\Psi}$ & $\frac{1}{24}N\left(N+1\right)\left(N+2\right)\left(N+3\right)$ & $\mathcal{O}\left(N^{4}\right)$\tabularnewline
\hline 
\end{tabular}
\end{table}

In addition, the non-parametric approach also poses tremendous challenges
in computing the objectives values, gradients, and the Hessian of
the third and fourth central moments for a given portfolio \cite{zhou2021solving}.
Here, we exhibit the corresponding complexities in Table \ref{tab:Computational-complexity-of}.
As a result, existing first-order methods could not be efficient as
they often require many iterations to converge while per-iteration
cost is very high. On the other hand, existing second-order methods
are not scalable because the complexity of computing $\nabla^{2}\phi_{4}\left(\mathbf{w}\right)$
is $\mathcal{O}\left(N^{5}\right)$. 

\begin{table}
\caption{Formulations and computational complexity of computing high-order
moments in non-parametric way. \label{tab:Computational-complexity-of}}

\centering{}%
\begin{tabular}{|c|c|c|c|}
\hline 
\multicolumn{2}{|c|}{} & Formulation & Complexity\tabularnewline
\hline 
3rd & $\phi_{3}\left(\mathbf{w}\right)$ & $\mathbf{w}^{T}\boldsymbol{\Phi}\left(\mathbf{w}\otimes\mathbf{w}\right)$ & $\mathcal{O}\left(N^{3}\right)$\tabularnewline
\cline{2-4} 
central & $\nabla\phi_{3}\left(\mathbf{w}\right)$ & $3\boldsymbol{\Phi}\left(\mathbf{w}\otimes\mathbf{w}\right)$ & $\mathcal{O}\left(N^{3}\right)$\tabularnewline
\cline{2-4} 
moment & $\nabla^{2}\phi_{3}\left(\mathbf{w}\right)$ & $6\boldsymbol{\Phi}\left(\mathbf{I}\otimes\mathbf{w}\right)$ & $\mathcal{O}\left(N^{4}\right)$\tabularnewline
\hline 
4th & $\phi_{4}\left(\mathbf{w}\right)$ & $\mathbf{w}^{T}\boldsymbol{\Psi}\left(\mathbf{w}\otimes\mathbf{w}\otimes\mathbf{w}\right)$ & $\mathcal{O}\left(N^{4}\right)$\tabularnewline
\cline{2-4} 
central & $\nabla\phi_{4}\left(\mathbf{w}\right)$ & $4\boldsymbol{\Psi}\left(\mathbf{w}\otimes\mathbf{w}\otimes\mathbf{w}\right)$ & $\mathcal{O}\left(N^{4}\right)$\tabularnewline
\cline{2-4} 
moment & $\nabla^{2}\phi_{4}\left(\mathbf{w}\right)$ & $12\boldsymbol{\Psi}\left(\mathbf{I}\otimes\mathbf{w}\otimes\mathbf{w}\right)$ & $\mathcal{O}\left(N^{5}\right)$\tabularnewline
\hline 
\end{tabular} 
\end{table}

Therefore, in the next section, we would present a parametric approach
to model the skewness and kurtosis such that the concerns discussed
above can be significantly eliminated. 

\section{Modeling high-order moments using generalized hyperbolic multivariate
skew-$t$ distribution \label{sec:Modeling-higher-order-using-MVST}}

In this section, we illustrate how to apply a parametric distribution
to model the data and derive the high-order moments from the parametric
model. To be more specific, this approach assumes that the assets'
returns follow a multivariate generalized hyperbolic skew-$t$ distribution.
Then, high-order moments can be represented using the parameters of
the fitted distribution. To proceed, we will first present some preliminary
knowledge of the generalized hyperbolic skew-$t$ distribution, followed
by the derivation of efficient methods for computing high-order moments
based on this distribution.

\subsection{ghMST Distribution }

The generalized hyperbolic multivariate skew-$t$ (ghMST) distribution
\cite{aas2006generalized,wei2019mixtures}, is a sub-class of the
generalized hyperbolic distribution \cite{barndorff1977exponentially},
which is often used in economics to model the data with skewness and
heavy tails \cite{hellmich2011efficient,hu2007risk,birge2016portfolio,haas2009financial}. 

Suppose that a $N$-dimensional random vector $\mathbf{x}$ follows
the ghMST distribution, i.e., $\mathbf{x}\sim\textsf{ghMST}\left(\boldsymbol{\mu},\boldsymbol{\Sigma},\boldsymbol{\gamma},\nu\right)$.
It has the probability density function (pdf) 
\begin{equation}
\begin{array}{rl}
f_{\textsf{ghMST}}\left(\mathbf{x}\left|\boldsymbol{\mu},\boldsymbol{\Sigma},\boldsymbol{\gamma},\nu\right.\right)=\frac{e^{\left(\mathbf{x}-\boldsymbol{\mu}\right)^{T}\boldsymbol{\Sigma}^{-1}\boldsymbol{\gamma}}}{\left(2\pi\right)^{\frac{N}{2}}\left|\boldsymbol{\Sigma}\right|{}^{\frac{1}{2}}}\cdot2\left(\frac{\nu}{2}\right)^{\frac{\nu}{2}}\cdot\frac{1}{\Gamma\left(\frac{\nu}{2}\right)}\cdot\quad\\
\text{ }\left(\frac{\chi+Q\left(\mathbf{x}\right)}{\boldsymbol{\gamma}^{T}\boldsymbol{\Sigma}^{-1}\boldsymbol{\gamma}}\right)^{-\frac{\nu+N}{4}}\cdot K_{-\frac{\nu+N}{2}}\left(\sqrt{\left(\nu+Q\left(\mathbf{x}\right)\right)\left(\boldsymbol{\gamma}^{T}\boldsymbol{\Sigma}^{-1}\boldsymbol{\gamma}\right)}\right),
\end{array}\label{eq:ghMST_pdf}
\end{equation}
where $\nu\in\mathbb{R}_{++}$ is the degree of freedom, $\boldsymbol{\mu}\in\mathbb{R}^{N}$
is the location vector, $\boldsymbol{\gamma}\in\mathbb{R}^{N}$ is
the skewness vector, $\boldsymbol{\Sigma}\in\mathbb{R}^{N\times N}$
is the scatter matrix, $\Gamma$ is the gamma function, $Q\left(\mathbf{x}\right)=\left(\mathbf{x}-\boldsymbol{\mu}\right)^{T}\boldsymbol{\Sigma}^{-1}\left(\mathbf{x}-\boldsymbol{\mu}\right)$,
and $K_{\lambda}$ is the modified Bessel function of the second kind
with index $\lambda$ \cite{barndorff2012levy}.
\begin{rem}
In the following contexts, $\boldsymbol{\mu}$ and $\boldsymbol{\Sigma}$
refer to the parameters of ghMST distribution, that is, to the location
vector and scatter matrix and not to the mean vector and covariance
matrix.
\end{rem}
Interestingly, the ghMST distribution can be represented in a hierarchical
structure as
\begin{equation}
\begin{aligned}\mathbf{x}|\tau\,\, & \overset{\text{i.i.d}}{\sim}\,\,\mathcal{N}\left(\boldsymbol{\mu}+\frac{1}{\tau}\boldsymbol{\gamma},\frac{1}{\tau}\boldsymbol{\Sigma}\right),\\
\tau\,\, & \overset{\text{i.i.d}}{\sim}\,\,\text{Gamma}\left(\frac{\nu}{2},\frac{\nu}{2}\right),
\end{aligned}
\label{hierarchical structure}
\end{equation}
 where $\mathcal{N}\left(\tilde{\boldsymbol{\mu}},\tilde{\boldsymbol{\Sigma}}\right)$
denotes the multivariate Gaussian distribution with mean vector $\tilde{\boldsymbol{\mu}}$
and covariance matrix $\tilde{\boldsymbol{\Sigma}}$, and $\text{Gamma}\left(a,b\right)$
represents the gamma distribution of shape $a$ and rate $b$.

\begin{figure}
\begin{centering}
\includegraphics[width=8.4cm]{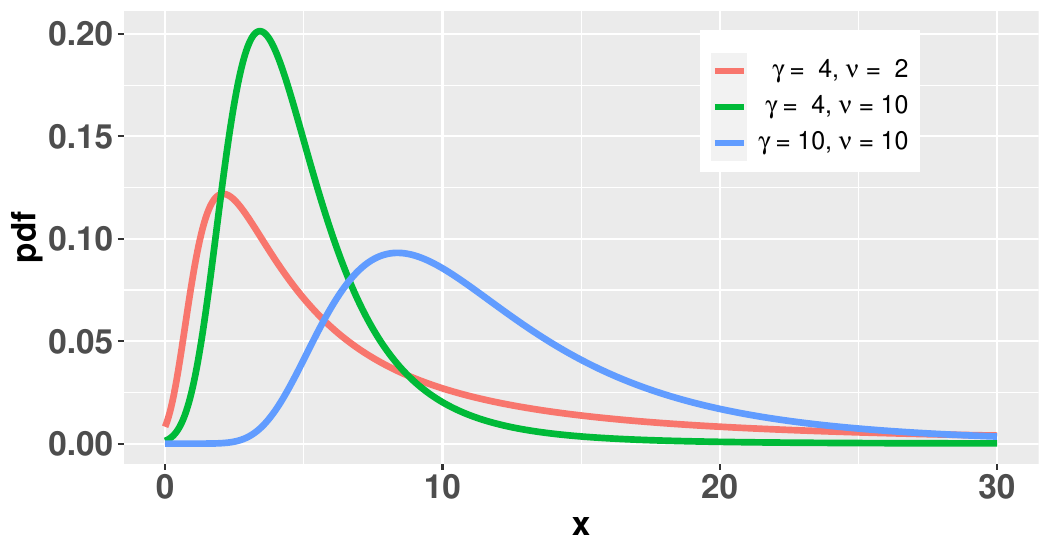}
\par\end{centering}
\caption{Illustrations for the univariate generalized hyperbolic skew-$t$
distribution $(\mu=0$, $\Sigma=1)$. \label{fig: skew-t-illustration}}
\end{figure}

Figure \ref{fig: skew-t-illustration} illustrates the skewness and
fat-tailness under the ghMST distribution. When $\boldsymbol{\gamma}$
is fixed, the higher the value of $\nu$, the thinner the tails. When
$\nu$ is fixed, the larger the value of $\boldsymbol{\gamma}$, the
heavier the skewness. Henceforth, the third- and fourth-moments are
naturally embedded into the parameters of the distribution. 

In the literature, some restricted multivariate skew-$t$ (rMST) distributions\footnote{Variants of rMST distribution include Gupta's skew-$t$ \cite{gupta2003multivariate},
Pyne\textquoteright s skew-$t$ \cite{pyne2009automated}, Branco\textquoteright s
skew-$t$ \cite{branco2001general}, and Azzalini\textquoteright s
skew-$t$ \cite{azzalini2003distributions}. It can be shown that
these variants have similar forms and can characterize the same distribution
after some parametrization \cite{lee2013mixtures}. } \cite{sahu2003new} are also capable of modeling asymmetry and fat-tailness.
In this paper, we choose to use the ghMST distribution for two reasons.
Foremost, the ghMST distribution is the only skew-$t$ distribution
that we can fit within a reasonable amount of time under high-dimensional
settings \cite{lee2013mixtures,lee2014finite}. The details of fitting
time are discussed in Appendix \ref{subsec:Comparing_estimation_time}.
In short, existing implementations\footnote{For fitting rMST distribution, we apply the EM algorithm \cite{wang2009multivariate}
implemented in R package\textcolor{black}{{} $\mathsf{EMMIXskew}$}
\cite{wang2018package}. } may take a number of minutes to fit the rMST distribution when $N\geq30$.
In contrast, existing EM algorithms can efficiently fit the ghMST
distribution \footnote{The ghMST distribution fitting process is carried out using the `fit\_mvst'
function from the R package\textcolor{black}{{} $\mathsf{fitHeavyTail}$}
\cite{palomarfitheavytail}.} with thousands of assets in few seconds \cite{aas2006generalized,breymann2013ghyp,mcneil2015quantitative}.
When $N=20$, fitting a ghMST distribution is over four orders of
magnitude faster than fitting an rMST distribution.

\begin{figure}
\begin{centering}
\includegraphics[width=8.8cm]{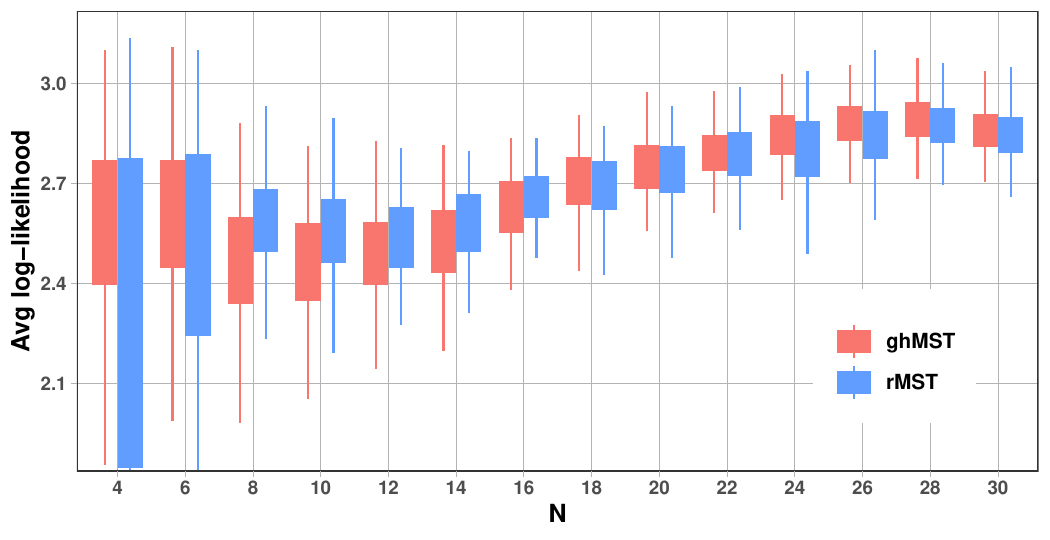}
\par\end{centering}
\caption{Out-of-sample log-likelihood for the restricted skew-t (rMST) and
generalized hyperbolic skew-$t$ (ghMST) distributions. \label{fig:Measuring-the-goodness-of-fit}}
\end{figure}

On the other hand, rMST distributions do not provide better out-of-sample
fitting performance. To show this, we conduct a simple experiment
as follows. In each realization, we randomly select $N$ assets from
SP$500$ stock list. Then, we randomly pick the data from $15N$ continuous
trading days to form the data set $\mathcal{D}$. Without shuffle,
$\mathcal{D}$ is split into training set $\mathcal{D}_{\text{train}}$
and test set $\mathcal{D}_{\text{test}}$ by assigning the $2/3$
data to the former and the remaining $1/3$ to the latter. For each
distribution, the optimal parameters are obtained via the training
set
\begin{equation}
\begin{array}{cc}
 & \boldsymbol{\Theta}^{\star}=\arg\max_{\boldsymbol{\Theta}}\text{ }\mathcal{L}\left(\mathcal{D}_{\text{train}};\boldsymbol{\Theta}\right).\end{array}\label{eq:boldsymboltheta}
\end{equation}
Then we compute the out-of-sample normalized log-likelihood on the
test set as $\frac{1}{5N^{2}}\mathcal{L}_{\text{test}}\left(\mathcal{D}_{\text{test}};\boldsymbol{\Theta}^{\star}\right).$
We repeat the experiments $50$ times for each problem dimension.
Figure \ref{fig:Measuring-the-goodness-of-fit} shows that the ghMST
distribution gives a higher average likelihood values when $N$ goes
large. As it is difficult to differentiate their obtained likelihoods,
the ghMST distribution appears to be the best choice for characterizing
high-order moments due to its significantly more efficient estimation.

\subsection{Computing high-order moments under ghMST distribution}

Incorporating the ghMST distribution into the design of high-order
portfolios also makes it convenient to manipulate the high-order moments,
i.e., skewness and kurtosis. In this subsection, we highlight two
advantages of using the parametric ghMST distribution. Firstly, it
allows for low-memory representation of the co-moments of the asset
return. Secondly, it provides more efficient computation of the skewness
and kurtosis to the portfolio returns. 

In the conventional framework, before we compute the high-order moments
of portfolio returns, we need to store large matrices, including $\boldsymbol{\Phi}$
and $\boldsymbol{\Psi}$. Now, we suppose that a random vector $\mathbf{r}$
follows a ghMST distribution. Then, according to Lemma \ref{lem:Skew_t_matrix},
the high-order moments can be easily computed from the parameter set
$\boldsymbol{\Theta}=\left\{ \boldsymbol{\mu},\boldsymbol{\Sigma},\boldsymbol{\gamma},\nu\right\} $,
which is significantly smaller than $\boldsymbol{\Phi}$ and $\boldsymbol{\Psi}$.
As a result, the memory complexity is reduced from $\mathcal{O}\left(N^{4}\right)$
to $\mathcal{O}\left(N^{2}\right)$. 
\begin{lem}
\label{lem:Skew_t_matrix}Assuming a random vector $\mathbf{r}\sim\textsf{ghMST}\left(\boldsymbol{\mu},\boldsymbol{\Sigma},\boldsymbol{\gamma},\nu\right)$,
then the mean and covariance of $\mathbf{r}$ are given as 
\begin{equation}
\begin{aligned}\mathbb{E}\left[\mathbf{r}\right] & =\boldsymbol{\mu}+a_{1}\boldsymbol{\gamma},\\
\text{Cov}\left[\mathbf{r}\right] & =a_{21}\boldsymbol{\Sigma}+a_{22}\boldsymbol{\gamma\gamma}^{T},
\end{aligned}
\label{eq:first_two_moments}
\end{equation}
where $a_{1}=\frac{\nu}{\nu-2}$, $a_{21}=\frac{\nu}{\nu-2}$, and
$a_{22}=\frac{2\nu^{2}}{\left(\nu-2\right)^{2}\left(\nu-4\right)}$
are scalar coefficients decided by $\nu$. The third moment co-skewness
matrix $\boldsymbol{\Phi}$ is expressed as 
\begin{align}
\Phi_{i,\left(j-1\right)\times N+k} & =a_{31}\gamma_{i}\gamma_{j}\gamma_{k}+\frac{a_{32}}{3}\left(\gamma_{i}\Sigma_{jk}+\gamma_{j}\Sigma_{ik}+\gamma_{k}\Sigma_{ij}\right).\label{eq:boldsymbolphi}
\end{align}
The fourth moment co-kurtosis matrix $\boldsymbol{\Psi}$ is expressed
as
\begin{align}
 & \Psi_{i,\left(j-1\right)N^{2}+\left(k-1\right)N+l}\nonumber \\
\, & =a_{41}\gamma_{i}\gamma_{j}\gamma_{k}\gamma_{l}+\frac{a_{42}}{6}\underbrace{\left(\Sigma_{ij}\gamma_{k}\gamma_{l}+\cdots+\Sigma_{kl}\gamma_{i}\gamma_{j}\right)}_{6\text{ items}}\nonumber \\
 & \qquad+\frac{a_{43}}{3}\left(\Sigma_{ij}\Sigma_{kl}+\Sigma_{ik}\Sigma_{jl}+\Sigma_{il}\Sigma_{jk}\right).\label{eq:sss}
\end{align}
 Here $a_{31}=\frac{16\nu^{3}}{\left(\nu-2\right)^{3}\left(\nu-4\right)\left(\nu-6\right)}$,
$a_{32}=\frac{6\nu^{2}}{\left(\nu-2\right)^{2}\left(\nu-4\right)}$,
$a_{41}=\frac{(12\nu+120)\nu^{4}}{(\nu-2)^{4}(\nu-4)(\nu-6)(\nu-8)}$,
$a_{42}=\frac{6(2\nu+4)\nu^{3}}{(\nu-2)^{3}(\nu-4)(\nu-6)}$ and $a_{43}=\frac{3\nu^{2}}{(\nu-2)(\nu-4)}$
are coefficients determined by $\nu$.
\end{lem}
\begin{IEEEproof}
See Appendix \ref{subsec:Proof-for-Lemma 1}.
\end{IEEEproof}
Another advantage of using the ghMST model comes at computing the
high-order central moments of the portfolio return in a fast way.
Specifically, though recovering the complete forms of $\boldsymbol{\Phi}$
and $\boldsymbol{\Psi}$ using Lemma \ref{lem:Skew_t_matrix} can
be computationally expensive, the skewness and kurtosis of $\mathbf{w}^{T}\mathbf{r}$
can be efficiently derived. 
\begin{lem}
\label{lem:portfolio_high_order_moments}Assuming $\mathbf{r}\sim\textsf{ghMST}\left(\boldsymbol{\mu},\boldsymbol{\Sigma},\boldsymbol{\gamma},\nu\right)$,
then the first-to-fourth central moments of $\mathbf{w}^{T}\mathbf{r}$,
denoted as $\phi_{i}\left(\mathbf{w}\right),i=1,\dots,4,$ are given
as follows
\begin{equation}
\begin{aligned}\phi_{1}\left(\mathbf{w}\right)=\, & \mathbf{w}^{T}\boldsymbol{\mu}+a_{1}\mathbf{w}^{T}\boldsymbol{\gamma},\\
\phi_{2}\left(\mathbf{w}\right)=\, & a_{21}\mathbf{w}^{T}\boldsymbol{\Sigma}\mathbf{w}+a_{22}\left(\mathbf{w}^{T}\boldsymbol{\gamma}\right)^{2},\\
\phi_{3}\left(\mathbf{w}\right)=\, & a_{31}\left(\mathbf{w}^{T}\boldsymbol{\gamma}\right)^{3}+a_{32}\left(\mathbf{w}^{T}\boldsymbol{\gamma}\right)\left(\mathbf{w}^{T}\boldsymbol{\Sigma}\mathbf{w}\right),\\
\phi_{4}\left(\mathbf{w}\right)=\, & a_{41}\left(\mathbf{w}^{T}\boldsymbol{\gamma}\right)^{4}+a_{42}\left(\mathbf{w}^{T}\boldsymbol{\gamma}\right)^{2}\left(\mathbf{w}^{T}\boldsymbol{\Sigma}\mathbf{w}\right)\\
 & \quad\quad+a_{43}\left(\mathbf{w}^{T}\boldsymbol{\Sigma}\mathbf{w}\right)^{2}.
\end{aligned}
\label{eq:phi1234}
\end{equation}
 
\end{lem}
\begin{IEEEproof}
See Appendix \ref{subsec:Proof-for-Lemma 2}.
\end{IEEEproof}
\begin{table}
\caption{Computational complexity of computing high-order moments using ghMST
distribution.\label{tab:Computational-complexity-of-2}}

\centering{}%
\begin{tabular}{|c|c|c|c|}
\hline 
 & Objective & Gradient & Hessian\tabularnewline
\hline 
\multirow{1}{*}{$3$-rd moment} & $\mathcal{O}\left(N^{2}\right)$ & $\mathcal{O}\left(N^{2}\right)$ & $\mathcal{O}\left(N^{3}\right)$\tabularnewline
\hline 
$4$-th moment & $\mathcal{O}\left(N^{2}\right)$ & $\mathcal{O}\left(N^{2}\right)$ & $\mathcal{O}\left(N^{3}\right)$\tabularnewline
\hline 
\end{tabular}
\end{table}

Under the ghMST distribution, we can significantly speed up the computation
of the objective value, gradient, and Hessian of high-order moments.
Their exact expressions are listed in Appendix \ref{subsec:Gradient-and-Hessian},
and their corresponding computational complexities are summarized
in Table \ref{tab:Computational-complexity-of-2}. Note that the per-iteration
complexity for first-order approaches has been reduced to $\mathcal{O}\left(N^{2}\right)$.
In response to this, in Section \ref{sec:Solving-MVSK-Portfolios},
we present an algorithm that mainly utilizes gradient information.
As a result, the proposed algorithm can exhibit superior scalability
over state-of-the-art methods. 

\section{Proposed methods for solving MVSK Portfolios \label{sec:Solving-MVSK-Portfolios}}

In this section, we explore new practical algorithms for solving Problem
(\ref{eq:MVSK}) under the ghMST distribution. The proposed method
iteratively minimizes the objective values via searching a fixed point
of a projected gradient mapping. The section is organized as follows.
We first recast the optimization problem (\ref{eq:MVSK}) as a fixed-point
problem. After that, we introduce a fixed-point acceleration scheme
to solve the fixed point more efficiently. To overcome the convergence
issues caused by the acceleration scheme, we further enhance the robustness
of the fixed-point acceleration method and accomplish our algorithm.
Finally, we provide an analysis of the complexity and convergence
of our proposed methods.

\subsection{Constructing the Fixed-point Problem \label{subsec:Equivalent-Fixed-point-Problem}}

Considering a continuous vector-to-vector mapping $G:\mathbb{R}^{N}\rightarrow\mathbb{R}^{N}$,
a point $\mathbf{w}$ is a fixed point of function $G$ when it satisfies
$\mathbf{w}=G\left(\mathbf{w}\right)$. In optimization, many iterative
methods aim at generating a sequence $\left\{ \mathbf{w}^{1},\mathbf{w}^{2},\dots\right\} $
that is expected to converge to a stationary point via a designed
update rule $\mathbf{w}^{k+1}=G\left(\mathbf{w}^{k}\right)$. As a
result, when those algorithms converge, the obtained $\mathbf{w}^{\star}$
is also the fixed point of $G$. In this subsection, we will introduce
the exact expression of $G$ of interest and how solving Problem (\ref{eq:MVSK})
can be transformed into finding a fixed point of function $G$.

The function $G$ we consider is selected as
\begin{equation}
G\left(\mathbf{w}^{k};\eta\right)\stackrel{\Delta}{=}\mathcal{P}_{\mathcal{W}}\left(\mathbf{w}^{k}-\eta\nabla f\left(\mathbf{w}^{k}\right)\right),\label{eq:AMM_update}
\end{equation}
where $\eta>0$ is the step size and the operator $\mathcal{P}_{\mathcal{W}}$
is defined as a projection onto a unit simplex \cite{condat2016fast}
\begin{equation}
\mathcal{P}_{\mathcal{W}}\left(\mathbf{w}^{k}\right)=\arg\min_{\mathbf{w}\in\mathcal{W}}\left\Vert \mathbf{w}-\mathbf{w}^{k}\right\Vert _{2}^{2},\label{eq:projection}
\end{equation}
which is a continuous vector-valued function defined on $\mathbf{w}\in\mathbb{R}^{N}$. 
\begin{rem}
In fact, the choice of $G$ is not unique, but (\ref{eq:AMM_update})
is preferred because it is simple to manipulate. Instead of calling
a quadratic programming solver, we can design a water-filling algorithm
\cite{palomar2005practical} to solve $G\left(\mathbf{w}^{k};\eta\right)$
efficiently. Details are elaborated in Section \ref{subsec:Waterfilling-algorithm}.
The simplicity of solving $G$ plays an important role in promoting
the efficiency and scalability of the the proposed algorithm. 
\end{rem}
Given any $\eta>0$, the fixed point of $G$ is the stationary point
of Problem (\ref{eq:MVSK}). This is shown in Lemma \ref{lem:The-set-of-fixed-point}.
\begin{lem}
\label{lem:The-set-of-fixed-point}The set of fixed point of $G\left(\cdot;\eta\right)$,
i.e., $\mathbf{w}=G\left(\mathbf{w};\eta\right)$, coincides with
that of the stationary points of Problem (\ref{eq:MVSK}).
\end{lem}
\begin{IEEEproof}
Let $\mathbf{w}^{\star}\in\mathcal{W}$ be the fixed point of $G\left(\cdot;\eta\right)$,
i.e., $\mathbf{w}^{\star}=G\left(\mathbf{w}^{\star};\eta\right)$.
Hence, $\mathbf{w}^{\star}$ is the optimal solution to the following
convex optimization problem:
\begin{equation}
\begin{array}{ll}
\underset{\mathbf{w}}{\mathsf{minimize}}\, & \frac{1}{2}\left\Vert \mathbf{w}-\left(\mathbf{w}^{\star}-\eta\nabla f\left(\mathbf{w}^{\star}\right)\right)\right\Vert _{2}^{2}\\
\mathsf{subject}\text{ }\mathsf{to} & \mathbf{w}\in\mathcal{W}.
\end{array}\label{eq:projection_optimal}
\end{equation}
 Therefore, for any $\mathbf{y}\in\mathcal{W}$, we have 
\begin{align}
 & \quad\left(\mathbf{y}-\mathbf{w}^{\star}\right)^{T}\left(\mathbf{w}^{\star}-\left(\mathbf{w}^{\star}-\eta\nabla f\left(\mathbf{w}^{\star}\right)\right)\right)\nonumber \\
 & =\eta\left(\mathbf{y}-\mathbf{w}^{\star}\right)^{T}\nabla f\left(\mathbf{w}^{\star}\right)\geq0,\label{eq:stationary}
\end{align}
which already indicates that $\mathbf{w}^{\star}$ is the stationary
point of Problem (\ref{eq:MVSK}).
\end{IEEEproof}
Using Lemma \ref{lem:The-set-of-fixed-point}, we can recast Problem
(\ref{eq:MVSK}) into the following optimization problem 
\begin{equation}
\text{\ensuremath{\mathsf{find}} }\mathbf{w}\in\mathcal{W},\text{ \ensuremath{\mathsf{subject}} \ensuremath{\mathsf{to}} }\mathbf{w}=G\left(\mathbf{w};\eta\right).\label{eq:fixed_problem}
\end{equation}
This well-known fixed-point problem can be solved by the fixed-point
iteration method \cite{judd1998numerical}, which iterates the following
update 
\begin{equation}
\mathbf{w}^{k+1}=G\left(\mathbf{w}^{k};\eta\right),\label{eq:conventional_update}
\end{equation}
in which the function $G$ should be Lipschitz continuous with Lipschitz
constant $L<1$. In practice, this conventional approach is often
criticized for slow convergence. Hence, in the rest part of this section,
we will introduce an acceleration scheme that significantly improves
its convergence.

\subsection{Fixed-point Acceleration \label{subsec:Fixed-point-Acceleration}}

We first reformulate the fixed-point problem as finding a root of
a residual function $R:\mathbb{R}^{N}\rightarrow\mathbb{R}^{N}$
\begin{equation}
R\left(\mathbf{w};\eta\right)=G\left(\mathbf{w};\eta\right)-\mathbf{w}.\label{eq:residual_function}
\end{equation}
If the problem is unconstrained, the non-smooth version of Newton-Raphson
method \cite{qi1993nonsmooth} solves the fixed-point problem via
iterating the following update formula 
\begin{equation}
\mathbf{w}^{k+1}=\mathbf{w}^{k}-\mathbf{M}^{-1}\left(\mathbf{w}^{k};\eta\right)R\left(\mathbf{w}^{k};\eta\right),\label{eq:Newton_update_original}
\end{equation}
where $\mathbf{M}\left(\mathbf{w}^{k};\eta\right)\in\mathbb{R}^{N\times N}\in\partial R\left(\mathbf{w}^{k};\eta\right)$
and $\partial R\left(\mathbf{w}^{k};\eta\right)$ is the Clarke's
generalized Jacobian of $R$ evaluated at $\mathbf{w}=\mathbf{w}^{k}$
\cite{clarke1990optimization}. However, (\ref{eq:Newton_update_original})
is not applicable in our case. On one hand, the acceleration may render
iterates infeasible, i.e., $\mathbf{w}^{k+1}\notin\mathcal{W}$. To
make up for it, a heuristic alternative to (\ref{eq:Newton_update_original})
is 
\begin{equation}
\mathbf{w}^{k+1}=\mathcal{P}_{\mathcal{W}}\left(\mathbf{w}^{k}-\mathbf{M}^{-1}\left(\mathbf{w}^{k};\eta\right)R\left(\mathbf{w}^{k};\eta\right)\right).\label{eq:Newton_update}
\end{equation}
On the other hand, $\mathbf{\mathbf{M}}\left(\mathbf{w}^{k};\eta\right)$
is generally intractable to obtain. But we notice that the classical
directional derivative evaluated at $\mathbf{w}=\mathbf{w}^{k}$ still
exists and is given as
\begin{equation}
D_{\mathbf{d}}R\left(\mathbf{w}^{k};\eta\right)=\lim_{h\rightarrow0}\frac{R\left(\mathbf{w}^{k}+h\mathbf{d};\eta\right)-R\left(\mathbf{w}^{k};\eta\right)}{h}.\label{eq:DD}
\end{equation}
Then, according to \cite[ Lemma 2.2]{qi1993nonsmooth}, for any direction
$\mathbf{d}$, there exists a matrix $\mathbf{M}\left(\mathbf{w}^{k};\eta\right)\in\partial R\left(\mathbf{w}^{k};\eta\right)$
such that 

\begin{equation}
D_{\mathbf{d}}R\left(\mathbf{w}^{k};\eta\right)=\mathbf{M}\left(\mathbf{w}^{k};\eta\right)\mathbf{d},\label{eq:dd}
\end{equation}
Hence, by assigning $h=1$ and $\mathbf{d}=G\left(\mathbf{w}^{k};\eta\right)-\mathbf{w}^{k}$
to (\ref{eq:DD}), we can construct the secant equation at $\mathbf{w}=\mathbf{w}^{k}$
as 

\begin{equation}
\mathbf{M}\left(\mathbf{w}^{k};\eta\right)R\left(\mathbf{w}^{k};\eta\right)\approx V\left(\mathbf{w}^{k};\eta\right),\label{eq:secant_equation}
\end{equation}
where the function $V:\mathbb{R}^{N}\rightarrow\mathbb{R}^{N}$ is
defined as
\begin{align}
V\left(\mathbf{w}^{k};\eta\right) & \stackrel{\Delta}{=}R\left(G\left(\mathbf{w}^{k};\eta\right);\eta\right)-R\left(\mathbf{w}^{k};\eta\right)\nonumber \\
 & =G\left(G\left(\mathbf{w}^{k};\eta\right);\eta\right)-2G\left(\mathbf{w}^{k};\eta\right)+\mathbf{w}^{k}.\label{eq:V(w)}
\end{align}
Here, we replace the matrix $\mathbf{M}\left(\mathbf{w}^{k};\eta\right)$
by the scaled identity matrix $\left(\alpha^{k}\right)^{-1}\mathbf{I}$
such that the inverse of it can be easily derived. The value of $\alpha^{k}$
is therefore determined by approximating the following equation
\begin{equation}
\left(\alpha^{k}\right)^{-1}R\left(\mathbf{w}^{k};\eta\right)\approx V\left(\mathbf{w}^{k};\eta\right),\label{eq:new_secant}
\end{equation}
whose details will be elaborated later. As a result, we have the formulation
for the first-level fixed-point acceleration, i.e., 
\begin{equation}
\mathbf{y}_{1}^{k}\stackrel{\Delta}{=}\mathbf{w}^{k}-\alpha^{k}R\left(\mathbf{w}^{k};\eta\right).\label{eq:y1k}
\end{equation}
Intuitively, as a replacement to (\ref{eq:Newton_update}), the projection
of the new point $\mathcal{P}_{\mathcal{W}}\left(\mathbf{y}_{1}^{k}\right)$
is expected to provide smaller residual values compared to $\mathbf{w}^{k}$. 

Inspired by the \emph{`squared extrapolation method'} \cite{varadhan2004squared},
we introduce the second-level acceleration by defining 
\begin{equation}
\mathbf{y}_{2}^{k}\stackrel{\Delta}{=}\mathbf{y}_{1}^{k}-\alpha^{k}R\left(\mathbf{y}_{1}^{k};\eta\right).\label{eq:y2k}
\end{equation}
This strategy, inspired by \cite{raydan2002relaxed}, can be seen
as taking two successive first-level acceleration using the same step
length. Interestingly, the value of $R\left(\mathbf{y}_{1}^{k};\eta\right)$
can be approximated by manipulating the secant equations. To be more
specific, we assign different values of $\mathbf{d}$ to construct
the secant equations. In (\ref{eq:secant_equation}), $\mathbf{d}$
is set to $\mathbf{d}_{1}=G\left(\mathbf{w}^{k};\eta\right)-\mathbf{w}^{k}$.
Now, we set 
\begin{align}
\mathbf{d}_{2} & =-\alpha^{k}\left[G\left(\mathbf{w}^{k};\eta\right)-\mathbf{w}^{k}\right]=-\alpha^{k}\mathbf{d}_{1}.\label{eq:d2}
\end{align}
This indicates that the approximation of $R\left(\mathbf{y}_{1}^{k};\eta\right)-R\left(\mathbf{w}^{k};\eta\right)$
can be obtained by multiplying a scaling factor $-\alpha^{k}$ to
$V\left(\mathbf{w}^{k};\eta\right)$, i.e.,
\begin{align}
R\left(\mathbf{y}_{1}^{k};\eta\right)-R\left(\mathbf{w}^{k};\eta\right) & \approx-\alpha^{k}V\left(\mathbf{w}^{k};\eta\right).\label{eq:R(y1k)}
\end{align}
Therefore, we obtain the closed-form approximation for $\mathbf{y}_{2}^{k}$
as
\begin{align}
\mathbf{y}_{2}^{k} & \stackrel{\Delta}{=}\mathbf{w}^{k}-\alpha^{k}R\left(\mathbf{w}^{k};\eta\right)-\alpha^{k}\left[R\left(\mathbf{w}^{k};\eta\right)-\alpha^{k}V\left(\mathbf{w}^{k};\eta\right)\right]\nonumber \\
 & =\mathbf{w}^{k}-2\alpha^{k}R\left(\mathbf{w}^{k};\eta\right)+\left(\alpha^{k}\right)^{2}V\left(\mathbf{w}^{k};\eta\right).\label{eq:define_y2k}
\end{align}
Eventually, the update for $\mathbf{w}$ is finalized as
\begin{equation}
\mathbf{w}^{k+1}=\mathcal{P}_{\mathcal{W}}\left(\mathbf{w}^{k}-2\alpha^{k}R\left(\mathbf{w}^{k};\eta\right)+\left(\alpha^{k}\right)^{2}V\left(\mathbf{w}^{k};\eta\right)\right).\label{eq:proposed_update}
\end{equation}

Now we introduce how to compute the value of $\alpha^{k}$. In the
literature, $\alpha^{k}$ is usually estimated by minimizing a discrepancy
measure based on the secant equation (\ref{eq:new_secant}). From
\cite{varadhan2008simple}, we select $\left\Vert R\left(\mathbf{w}^{k};\eta\right)-\alpha V\left(\mathbf{w}^{k};\eta\right)\right\Vert ^{2}\big/\left|\alpha\right|$
as our discrepancy measure. In addition, because the term $R\left(\mathbf{w}^{k};\eta\right)$
in (\ref{eq:y1k}) can be seen as a direction to achieve small objective
values, it is naturally to impose the constraint $\alpha^{k}\leq0$
such that the acceleration is performed along with descent direction. 

Meanwhile, we require another constraint 
\begin{equation}
\left\langle \mathbf{y}_{1}^{k}-\mathbf{w}^{k},\mathbf{y}_{2}^{k}-\mathbf{y}_{1}^{k}\right\rangle \geq0.\label{eq:critical__iniequality}
\end{equation}
We hope that the direction of first-level acceleration should be similar
to the direction of second-level acceleration. The inequality (\ref{eq:critical__iniequality}),
which is equivalent to

\begin{equation}
\left\langle R\left(\mathbf{w}^{k};\eta\right),R\left(\mathbf{w}^{k};\eta\right)-\alpha^{k}V\left(\mathbf{w}^{k};\eta\right)\right\rangle \geq0,\label{eq:ect}
\end{equation}
provides another constraint for the value of $\alpha^{k}$, i.e.,
$\alpha^{k}\geq b\left(\mathbf{w}^{k}\right)$, where the function
$b:\mathbb{R}^{N}\rightarrow\mathbb{R}$ is denoted as 
\begin{align}
 & b\left(\mathbf{w}^{k}\right)\nonumber \\
 & =\begin{cases}
\frac{\left\Vert R\left(\mathbf{w}^{k};\eta\right)\right\Vert _{2}^{2}}{\left\langle R\left(\mathbf{w}^{k};\eta\right),V\left(\mathbf{w}^{k};\eta\right)\right\rangle } & \text{if }\text{\ensuremath{\left\langle R\left(\mathbf{w}^{k};\eta\right),V\left(\mathbf{w}^{k};\eta\right)\right\rangle }}<0,\\
-\infty & \text{otherwise}.
\end{cases}\label{eq:bwk}
\end{align}

Therefore, the value of $\alpha^{k}$ is computed as the solution
to the following constrained least square problem
\begin{equation}
\begin{array}{ll}
\underset{\alpha}{\mathsf{minimize}}\, & \left\Vert R\left(\mathbf{w}^{k};\eta\right)-\alpha V\left(\mathbf{w}^{k};\eta\right)\right\Vert ^{2}\big/\left|\alpha\right|\\
\mathsf{subject}\text{ }\mathsf{to} & b\left(\mathbf{w}^{k}\right)\leq\alpha<0,
\end{array}\label{eq:alpha_problem}
\end{equation}
 whose solution can be easily obtained as
\begin{equation}
\alpha^{k}=\max\left(-\left\Vert R\left(\mathbf{w}^{k};\eta\right)\right\Vert \big/\left\Vert V\left(\mathbf{w}^{k};\eta\right)\right\Vert ,b\left(\mathbf{w}^{k}\right)\right).\label{eq:alphak}
\end{equation}

In principle, we can also simulate $\mathbf{y}_{i+1}^{k}\stackrel{\Delta}{=}\mathbf{y}_{i}^{k}-\alpha_{k}R\left(\mathbf{y}_{i}^{k};\eta\right)$
for $i>2$, but the formulations are typically more complicated to
derive and more levels of approximation is more likely to produce
invalid acceleration. 

Compared to the conventional update (\ref{eq:conventional_update}),
the proposed method only includes some small extra computational costs
at each iteration, while significantly improve the efficiency in practice.
However, like many other fixed point acceleration methods, directly
iterating (\ref{eq:proposed_update}) may not yield robust results.
In other words, we may obtain a sequence that does not converge. Hence,
we will provide our solutions to further improve the robustness of
the proposed fixed-point acceleration. 

\subsection{A Robust Fixed Point Acceleration (RFPA) Algorithm \label{subsec:Proposed-Algorithms}}

To establish a stable convergence, we require that the sequence $\left\{ f\left(\mathbf{w}^{k}\right)\right\} $
should be monotone, i.e.,
\begin{equation}
\forall k:f\left(\mathbf{w}^{k+1}\right)\leq f\left(\mathbf{w}^{k}\right).\label{eq:monotone_condition}
\end{equation}
The strategy is illustrated as follows. When the fixed-point acceleration
fails to improve the objective, i.e., $f\left(\mathbf{w}^{k+1}\right)>f\left(\mathbf{w}^{k}\right)$,
we first set $\mathbf{w}^{k+1}=G\left(\mathbf{w}^{k};\eta\right)$.
Then, we keep decreasing it by $\eta\leftarrow\beta\eta$ with a scaling
factor $\beta\in\left(0,1\right)$ until the following condition is
met
\begin{equation}
\begin{aligned}f\left(\mathbf{w}^{k+1}\right) & \leq f\left(\mathbf{w}^{k}\right)+\nabla f\left(\mathbf{w}^{k}\right)^{T}\left(\mathbf{w}^{k+1}-\mathbf{w}^{k}\right)\\
 & \quad+\frac{1}{2\eta}\left\Vert \mathbf{w}^{k}-\mathbf{w}^{k+1}\right\Vert _{2}^{2}.
\end{aligned}
\label{eq:line_search}
\end{equation}
Once the condition (\ref{eq:line_search}) holds, the sequence $\left\{ f\left(\mathbf{w}^{k}\right)\right\} $
is then monotone with the details provided in Appendix \ref{subsec:Monotonicity-of-the}.
Eventually, we summarize the proposed robust fixed point acceleration
(RFPA) algorithm in Algorithm \ref{alg: QN-MM for mvsk portfolio-1}. 

\begin{algorithm}
\caption{Robust Fixed Point Acceleration (RFPA) algorithm for solving Problem
(\ref{eq:MVSK}). \label{alg: QN-MM for mvsk portfolio-1}}
\begin{algorithmic}[1]

\STATE Initialize $\mathbf{w}^{0}\in\mathcal{W}$, $\eta$, $\eta_{0}$,
$\beta$

\FOR{$k=0,1,2,\ldots$} 

\STATE Compute $R\left(\mathbf{w}^{k};\eta\right)$, $V\left(\mathbf{w}^{k};\eta\right)$

\STATE $\alpha^{k}=\max\left(-\left\Vert R\left(\mathbf{w}^{k};\eta\right)\right\Vert \big/\left\Vert V\left(\mathbf{w}^{k};\eta\right)\right\Vert ,b\left(\mathbf{w}^{k}\right)\right)$.

\STATE $\mathbf{w}^{k+1}=$

$\quad\quad\mathcal{P}_{\mathcal{W}}\left(\mathbf{w}^{k}-2\alpha^{k}R\left(\mathbf{w}^{k};\eta\right)+\left(\alpha^{k}\right)^{2}V\left(\mathbf{w}^{k};\eta\right)\right)$.

\IF { $f\left(\mathbf{w}^{k+1}\right)>f\left(\mathbf{w}^{k}\right)$
} 

\STATE \textbf{$\eta'=\eta_{0}$.}

\STATE Update $\mathbf{w}^{k+1}=G\left(\mathbf{w}^{k};\eta'\right)$.

\WHILE {(\ref{eq:line_search}) not satisfied}

\STATE $\eta'\leftarrow\beta\eta'$, go to step 8.

\ENDWHILE

\ENDIF

\STATE Terminate loop if converges.

\ENDFOR

\end{algorithmic}
\end{algorithm}

If no fixed point acceleration is applied, we only iterate $\mathbf{w}^{k+1}=G\left(\mathbf{w}^{k};\eta\right)$
that satisfies (\ref{eq:line_search}), the RFPA algorithm would reduce
to the projected gradient descent (PGD) method. 

The main motivation of executing projected gradients is to enlarge
the difference between $\mathbf{w}^{k+1}$ and $\mathbf{w}^{k}$.
Theoretically, whether the fixed-point acceleration would significantly
improve the convergence is decided by the numerical properties at
$\mathbf{w}^{k}$. Therefore, if the difference of $\mathbf{w}^{k}$
and $\mathbf{w}^{k+1}$ is not large enough while the fixed-point
acceleration at $\mathbf{w}^{k}$ is not successful, the algorithm
tends to reject the fixed-point acceleration at $\mathbf{w}^{k+1}$
due to their similar numerical properties. 

\subsection{Complexity Analysis and Convergence Analysis \label{subsec:Complexity-Analysis}}

The overall complexity of the proposed RFPA algorithm is $\mathcal{O}\left(N^{2}\right)$.
Specifically, the per-iteration cost of the proposed RFPA algorithm
comes from two parts: computing the gradient $\nabla f\left(\mathbf{w}^{k}\right)$
and solving a projection problem $\mathcal{P}_{\mathbf{\mathcal{W}}}$.
With the help of the parametric skew-$t$ distribution, the computational
complexity of computing the gradient is reduced to $\mathcal{O}\left(N^{2}\right)$.
For solving the projection problems, the computational complexity
mainly depends on finding proper values of the dual variables via
bisection. According to the Section \ref{subsec:Waterfilling-algorithm}
of the Appendix, the primary cost of the water-filling algorithm is
to sort an array of numbers. Therefore, the corresponding complexity
is $\mathcal{O}\left(N\log N\right)$. In conclusion, regardless of
the number of outer iterations, the overall complexity of the proposed
RFPA algorithm is $\mathcal{O}\left(N^{2}\right)$. 

On the contrary, if we apply the non-parametric modeling of the high-order
moments, then the bottleneck of all the algorithms would be the computation
of the gradient or the Hessian, which are $\mathcal{O}\left(N^{4}\right)$
or $\mathcal{O}\left(N^{5}\right)$, respectively. After we assume
the returns follow a parametric skew-$t$ distribution, the complexity
of the second-order methods, like Q-MVSK algorithm and sequential
quadratic programming method, becomes $\mathcal{O}\left(N^{3}\right)$
due to the complexity of evaluating $\nabla^{2}\phi_{4}\left(\mathbf{w}\right)$. 

The convergence of the RFPA algorithm for MVSK portfolio optimization
is given as Theorem \ref{thm:If-,-then}. By solving the fixed point
of function $G$, we can obtain the stationary point of Problem (\ref{eq:MVSK}).
\begin{thm}
If $\mathbf{w}^{k}=\mathbf{w}^{k+1}$, then $\mathbf{w}^{k}$ is a
stationary point of Problem (\ref{eq:MVSK}). \label{thm:If-,-then}
\end{thm}
\begin{IEEEproof}
See Appendix \ref{subsec:Proof-of-Theorem 1}.
\end{IEEEproof}
Theorem \ref{thm:If-,-then} indicates that the algorithm can obtain
the stationary point of Problem (\ref{eq:MVSK}) if it terminates
with $\mathbf{w}^{k}=\mathbf{w}^{k+1}$, which always holds in empirical
studies as shown in Section \ref{subsec:Empirical-Convergence}.

\section{Extension: Solving MVSK-Tilting Portfolios with General Deterioration
Measure \label{sec:Solving-MVSK-Tilting-Portfolios}}

Our proposed framework provides an efficient and scalable discipline
for handling high-order moments, therefore presents great potential
for more advanced and sophisticated applications, like multi-period
portfolio optimization problems \cite{brandt2005simulation,cong2016multi},
incorporating diversification into the high-order designs \cite{mercurio2020entropy,kang2021entropy},
and increasing the robustness of current MVSK formulation \cite{chen2018robust}.
In this section, we explore an interesting example of extending our
framework to other portfolios.

In portfolio theory, though the MVSK framework finds a solution on
the efficient frontiers, choosing proper values for $\boldsymbol{\lambda}$
may be difficult and the optimal weights are often concentrated into
some positions, resulting in a greater idiosyncratic risk \cite{prakash2003selecting}.
Therefore, we can generalize the idea of the RFPA algorithm for solving
another important high-order portfolio called the MVSK-Tilting problem
with general deterioration measures. This MVSK-Tilting portfolio aims
at improving a given portfolio that is not sufficiently optimal from
the MVSK perspective by tilting it toward a direction that concurrently
ameliorates all the objectives \cite{jurczenko2006multi,boudt2020algorithmic}.

The problem of interest is formulated as 
\begin{equation}
\begin{array}{ll}
\underset{\mathbf{w},\delta}{\mathsf{minimize}}\,\,\, & -\delta+\lambda\cdot g_{\text{det}}\left(\mathbf{w}\right)\\
\mathsf{subject}\text{ }\mathsf{to\,}\,\, & \phi_{1}\left(\mathbf{w}\right)\geq\phi_{1}\left(\mathbf{w}_{0}\right)+d_{1}\delta,\\
 & \phi_{2}\left(\mathbf{w}\right)\leq\phi_{2}\left(\mathbf{w}_{0}\right)-d_{2}\delta,\\
 & \phi_{3}\left(\mathbf{w}\right)\geq\phi_{3}\left(\mathbf{w}_{0}\right)+d_{3}\delta,\\
 & \phi_{4}\left(\mathbf{w}\right)\leq\phi_{4}\left(\mathbf{w}_{0}\right)-d_{2}\delta,\\
 & \mathbf{w}\in\mathcal{W},
\end{array}\label{eq:MVSKTG}
\end{equation}
 where $\mathbf{d}=\left[\begin{array}{cccc}
d_{1} & d_{2} & d_{3} & d_{4}\end{array}\right]^{T}\geq\mathbf{0}$ represents the relative importance of each target, $g_{\text{det}}\left(\mathbf{w}\right)$
is a differentiable function that corresponds to an assigned deterioration
measure with respect to $\mathbf{w}_{0}$, and $\lambda$ is the regularization
coefficient. For example, $g_{\text{det}}\left(\mathbf{w}\right)$
can represent a tracking error 
\begin{equation}
g_{\text{det}}\left(\mathbf{w}\right)=\left(\mathbf{w}-\mathbf{w}_{0}\right)^{T}\text{Cov}\left[\mathbf{r}\right]\left(\mathbf{w}-\mathbf{w}_{0}\right).\label{eq:gdet}
\end{equation}
Implicitly, the point $\mathbf{w}_{0}$ refers to a reference portfolio
that satisfies $\mathbf{w}_{0}=\arg\min_{\mathbf{w}\in\mathcal{W}}g_{\text{det}}\left(\mathbf{w}\right)$,
indicating that the penalty would be imposed when we tilt $\mathbf{w}$
away from $\mathbf{w}_{0}$. 

As the key for the success of the RFPA algorithm is to form a separable
function $G$ such that the fixed point of $G$ is the stationary
point we want to obtain. The function $G$ corresponds to an optimization
problem that has the following properties:
\begin{itemize}
\item The objective function of the optimization problem is separable.
\item The constraint of the optimization problem is simple. In our case,
we require that the constraint is just $\mathbf{w}\in\mathcal{W}$.
\end{itemize}
Therefore, we first move the MVSK-Tilting constraints into the objective,
resulting in the following equivalent problem:
\begin{equation}
\begin{array}{ll}
\underset{\mathbf{w}}{\mathsf{minimize}} & \max\left[\varphi\left(\mathbf{w}\right)\right]+\lambda\cdot g_{\text{det}}\left(\mathbf{w}\right)\\
\mathsf{subject}\text{ }\mathsf{to} & \mathbf{w}\in\mathcal{W},
\end{array}\label{eq:Equivalent-MVSKT}
\end{equation}
 in which 
\begin{equation}
\varphi\left(\mathbf{w}\right)=\left[\begin{array}{c}
\varphi_{1}\left(\mathbf{w}\right)\\
\varphi_{2}\left(\mathbf{w}\right)\\
\varphi_{3}\left(\mathbf{w}\right)\\
\varphi_{4}\left(\mathbf{w}\right)
\end{array}\right]=\left[\begin{array}{c}
\frac{1}{d_{1}}\left[\phi_{1}\left(\mathbf{w}_{0}\right)-\phi_{1}\left(\mathbf{w}\right)\right]\\
\frac{1}{d_{2}}\left[\phi_{2}\left(\mathbf{w}\right)-\phi_{2}\left(\mathbf{w}_{0}\right)\right]\\
\frac{1}{d_{3}}\left[\phi_{3}\left(\mathbf{w}_{0}\right)-\phi_{3}\left(\mathbf{w}\right)\right]\\
\frac{1}{d_{4}}\left[\phi_{4}\left(\mathbf{w}\right)-\phi_{4}\left(\mathbf{w}_{0}\right)\right]
\end{array}\right].\label{eq:psi(w)}
\end{equation}

To alleviate the difficulty taken by the non-smoothness of the max
term, instead of directly solving Problem (\ref{eq:Equivalent-MVSKT}),
we solve the relaxation of (\ref{eq:Equivalent-MVSKT}) via the $\ell_{p}$-norm
smoothing approximation, i.e.,
\begin{equation}
\begin{array}{ll}
\underset{\mathbf{w}}{\mathsf{minimize}} & g_{p}\left(\mathbf{w}\right)=\left\Vert t\mathbf{1}+\varphi\left(\mathbf{w}\right)\right\Vert _{p}+\lambda\cdot g_{\text{det}}\left(\mathbf{w}\right)\\
\mathsf{subject}\text{ }\mathsf{to} & \mathbf{w}\in\mathcal{W},
\end{array}\label{eq:relaxed-MVSKT}
\end{equation}
 where $p$ is a positive integer, and $t$ is larger than any possible
value of the elements of $\varphi\left(\mathbf{w}\right)$ such that
\begin{equation}
\lim_{p\rightarrow\infty}\left\Vert t\mathbf{1}+\varphi\left(\mathbf{w}\right)\right\Vert _{p}-t=\max\left[\varphi\left(\mathbf{w}\right)\right].\label{eq:t}
\end{equation}
When the value of $p$ is large enough, the relaxed problem reduces
to the original problem. As $g_{p}\left(\mathbf{w}\right)$ is smooth,
the gradient exists for any $\mathbf{w}\in\mathcal{W}$, we have 
\begin{align}
 & \frac{\partial}{\partial\mathbf{w}}\left(\left\Vert t\mathbf{1}+\varphi\left(\mathbf{w}\right)\right\Vert _{p}\right)\nonumber \\
 & \quad=\left(\frac{\left(t\mathbf{1}+\varphi\left(\mathbf{w}\right)\right)^{T}}{\left\Vert t\mathbf{1}+\varphi\left(\mathbf{w}\right)\right\Vert _{p}}\right)^{p-1}\left[\begin{array}{c}
-\frac{1}{d_{1}}\nabla\phi_{1}\left(\mathbf{\mathbf{w}}\right)^{T}\\
\frac{1}{d_{2}}\nabla\phi_{2}\left(\mathbf{w}\right)^{T}\\
-\frac{1}{d_{3}}\nabla\phi_{3}\left(\mathbf{\mathbf{w}}\right)^{T}\\
\frac{1}{d_{4}}\nabla\phi_{4}\left(\mathbf{w}\right)^{T}
\end{array}\right].\label{eq:gradient_of_gp}
\end{align}
Hence, the relaxed problem is equivalent to find the fixed point of
the following function
\begin{equation}
G\left(\mathbf{w}^{k};\eta\right)\stackrel{\Delta}{=}\mathcal{P}_{\mathcal{W}}\left(\mathbf{w}^{k}-\eta\nabla g_{p}\left(\mathbf{w}^{k}\right)\right),\label{eq:GWk}
\end{equation}
where $\eta$ is the step size and 
\begin{align}
\nabla g_{p}\left(\mathbf{w}^{k}\right) & =\left.\frac{\partial}{\partial\mathbf{w}}\left(\left\Vert t\mathbf{1}+\varphi\left(\mathbf{w}\right)\right\Vert _{p}\right)\right|_{\mathbf{w}=\mathbf{w}^{k}}\nonumber \\
 & \quad+\lambda\left.\frac{\partial}{\partial\mathbf{w}}g_{\text{det}}\left(\mathbf{w}\right)\right|_{\mathbf{w}=\mathbf{w}^{k}}.\label{eq:nablagwk}
\end{align}
By simply applying Algorithm \ref{alg: QN-MM for mvsk portfolio-1},
the RFPA algorithm for the MVSK-Tilting problem with general deterioration
measure can be easily solved. 

\section{Numerical Simulations \label{sec:Numerical-Simulations}}

In this section, we conduct numerical experiments for evaluating our
proposed high-order portfolio solving framework\footnote{\textcolor{black}{We have released an R package $\mathsf{highOrderPortfolios}$
implementing our proposed algorithms at \href{https://github.com/dppalomar/highOrderPortfolios}{https://github.com/dppalomar/highOrderPortfolios}.}}. 

\subsection{On Applying the ghMST distribution }

The portfolios based on parametric representation of the high-order
moments distinguishes the portfolio obtained from traditional MVSK
framework. In other words, given the same data and optimization problem,
we can either compute $\phi_{i}\left(\mathbf{w}\right)$, $i=1,2,3,4,$
using non-parametric sample moments $\boldsymbol{\Phi}$ and $\boldsymbol{\Psi}$
in (\ref{eq:phianspsi}), or the parametric $\boldsymbol{\Theta}$
from ghMST distribution in Lemma \ref{lem:portfolio_high_order_moments},
resulting in different optimal portfolios. 

Assuming the data follows a ghMST distribution with the true parameter
$\boldsymbol{\Theta}_{\text{true}}$. We generate the synthetic data
set $\mathcal{D}$ based on $\boldsymbol{\Theta}_{\text{true}}$,
then construct the high-order portfolios using either non-parametric
approach or parametric skew-$t$ approach. Here we consider an MVSK
formulations with $\boldsymbol{\lambda}=\left(1,1,1,1\right)$ with
$\mathbf{w}_{\text{true}}$ as its optimal portfolio, i.e., 
\begin{align}
\mathbf{w}_{\text{true}} & =\arg\min_{\mathbf{w}\in\mathcal{\mathcal{W}}}f\left(\mathbf{w};\boldsymbol{\Theta}_{\text{true}},\boldsymbol{\lambda}\right).\label{eq:w_true}
\end{align}
Using the non-parametric approach, we first estimate $\boldsymbol{\Phi}$
and $\boldsymbol{\Psi}$ from $\mathcal{D}$, then obtain the optimal
portfolio $\mathbf{w}_{\text{np}}$ as the solution to (\ref{eq:MVSK}).
While with the parametric approach, we have to fit the ghMST distribution
given $\mathcal{D}$, then solve the optimal portfolio $\mathbf{w}_{\text{st}}$
based on the estimated parameters $\boldsymbol{\Theta}$. Here, we
denote the errors $\epsilon_{\text{np}}$ and $\epsilon_{\text{st}}$
as $\epsilon_{\text{np}}=\left\Vert \mathbf{w}_{\text{np}}-\mathbf{w}_{\text{true}}\right\Vert ^{2}$
and $\epsilon_{\text{st}}=\left\Vert \mathbf{w}_{\text{st}}-\mathbf{w}_{\text{true}}\right\Vert ^{2}$,
respectively. 

\begin{figure}
\begin{centering}
\includegraphics[width=8.5cm]{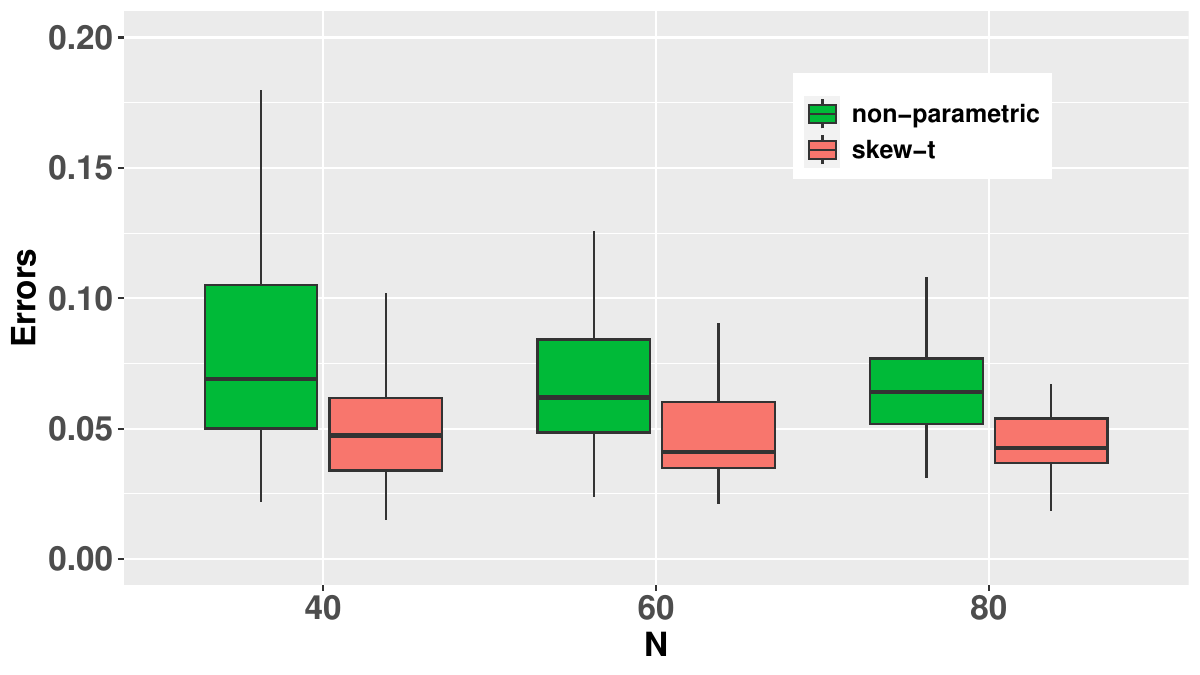}
\par\end{centering}
\caption{Errors of non-parametric and parametric approaches.\label{fig:Errors-using-non-parametric} }
\end{figure}

We repetitively evaluate the errors from different data sets under
different problem sizes. According to the result shown in the Figure
\ref{fig:Errors-using-non-parametric}, the parametric skew-$t$ approach
produces smaller errors than the non-parametric approach on any problem
size. 

\subsection{On Solving MVSK Portfolio Using RFPA Algorithm}

In this subsection, we conduct experiments to evaluate how applying
the ghMST distribution would accelerate the existing and proposed
algorithms and the performance of our proposed RFPA algorithm on efficiency
and scalability. We mainly utilize real-world data for the experiments.
The data is randomly selected from the S\&P 500 stock index. The trading
period is chosen from 2011-01-01 to 2020-12-31. 

\subsubsection{Comparing non-parametric and parametric (ghMST) approach}

We first perform the comparison on the non-parametric and parametric
modeling of the high-order moments. Given the data, we first estimate
the parameter $\boldsymbol{\Theta}$ for the ghMST distribution, then
generate the sample moments, i.e., sample skewness matrix and kurtosis
matrix, using Lemma \ref{lem:Skew_t_matrix}. In this way, $\phi_{i}\left(\mathbf{w}\right)$,
$i=1,2,3,4,$ will produce the same values under both non-parametric
and parametric modeling. 

We list the benchmarks as (first-order) MM algorithm \cite{zhou2021solving},
projected gradient descent (PGD) method, Q-MVSK (second-order SCA)
algorithm \cite{zhou2021solving}, the nonlinear optimization solver
`Nlopt' \cite{johnson2014nlopt} and our proposed RFPA algorithm.
The inner solver for QP is selected as $\textsf{quadprog}$ \cite{turlach2020quadprog}.
The weights $\boldsymbol{\lambda}$ are determined according to the
Constant Relative Risk Aversion utility function
\begin{equation}
\boldsymbol{\lambda}^{T}=\left[1,\text{ }\frac{\xi}{2},\text{ }\frac{\xi\left(\xi+1\right)}{6},\text{ }\frac{\xi\left(\xi+1\right)\left(\xi+2\right)}{24}\right],\label{eq:lambda}
\end{equation}
where $\xi\geq0$ is a parameter to measure the risk aversion \cite{boudt2015higher}.
Suggested by \cite{elminejad2022relative,barsky1997preference,pennacchi2008theory},
we set $\xi=6$ in this experiment. We further choose $\eta=5$, $\beta=0.5$
and investigate the empirical convergence of all algorithms under
two different dimensions $N=100$ and $N=400$. The gap is defined
as the difference of the objective value at each iteration and the
smallest objective value we obtained across all the methods. When
$N=400$, we cannot compare the performance of the non-parametric
approaches due to the memory limit that renders them intractable. 

\begin{figure}
\begin{centering}
\includegraphics[width=8.8cm]{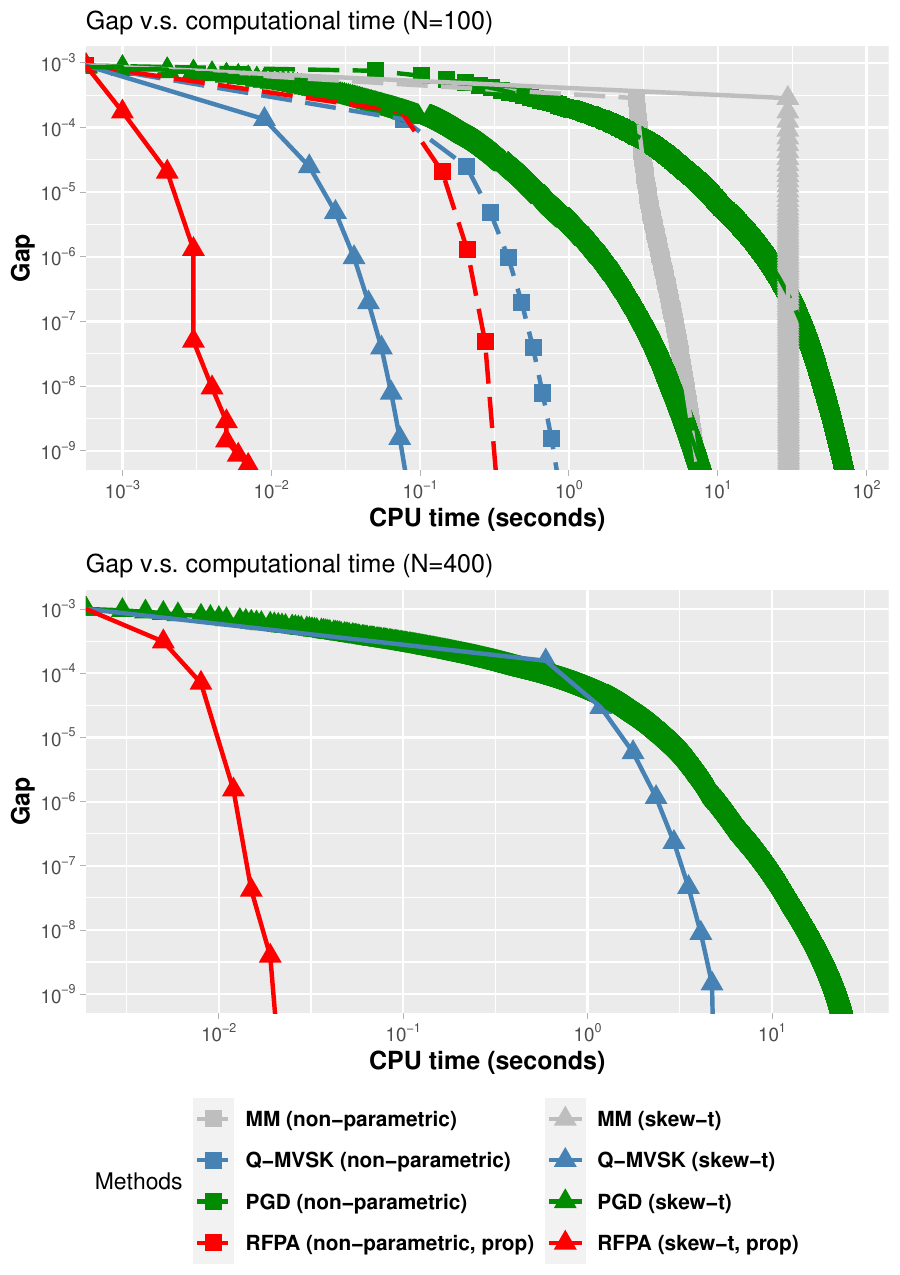}
\par\end{centering}
\caption{Convergence of algorithms for solving the MVSK portfolio optimization
problems (\ref{eq:MVSK}). \label{fig:Convergence-of-algorithms-on-MVSK}}

\end{figure}
We have the following observations according to the simulation results
exhibited in Figure \ref{fig:Convergence-of-algorithms-on-MVSK}.
When $N=100$, the time cost for Nlopt (Non-parametric) and Nlopt
(skew-$t$) is $9.349$ and $1.275$ seconds, respectively. When $N=400$,
the number for Nlopt(skew-$t$) becomes $52.934$ seconds. Note that
all the non-parametric approaches, which model the high-order moments
using sample moments, are not applicable in high dimension due to
the memory limit. Besides, MM methods require computing $\eta$ that
meets the condition $\frac{1}{\eta}\geq\sup_{\mathbf{w}\in\mathcal{W}}\left\Vert \nabla f\left(\mathbf{w}\right)\right\Vert _{2}$,
which is computationally expensive to obtain in high dimensional problems.
From the numerical simulations we observe the following.
\begin{itemize}
\item By applying the parametric skew-$t$ distribution, we can accelerate
the MVSK portfolio design by one-to-two orders of magnitude given
any optimization algorithm when $N=100$. 
\item The per-iteration cost of proposed RFPA and PGD algorithms is significantly
smaller than other methods with the help of water-filling algorithms. 
\item The effect of using the parametric skew-$t$ distribution tends to
be algorithm-dependent. The acceleration is more noticeable for first-order
algorithms like RFPA, which has negligible per-iteration cost. 
\end{itemize}

\subsubsection{Comparison on efficiency}

To better compare the efficiency of the proposed algorithms, we also
conduct experiments using real-world data sets with different problem
dimensions. For each problem size, we set $\eta=5$, $\beta=0.5$,
and take 200 independent experiments with $\xi$ randomly drawn from
the interval $\left(10^{-1},10\right)$. All the methods are initialized
with the same starting point $\mathbf{w}^{0}$. For Nlopt, the stopping
criteria are set as the default. For Q-MVSK, PGD, and RFPA, the algorithms
are regarded as converged when both the following conditions are satisfied:
\begin{equation}
\left|\mathbf{w}^{k+1}-\mathbf{w}^{k}\right|\leq10^{-6}\left(\left|\mathbf{w}^{k+1}\right|+\left|\mathbf{w}^{k}\right|\right),\label{eq:stoppingcriteria}
\end{equation}
\begin{align}
\left|f\left(\mathbf{w}^{k+1}\right)-f\left(\mathbf{w}^{k}\right)\right| & \leq10^{-6}\left(\left|f\left(\mathbf{w}^{k+1}\right)\right|+\left|f\left(\mathbf{w}^{k}\right)\right|\right).\label{eq:stoppingcondition}
\end{align}

\begin{figure}
\begin{centering}
\includegraphics[width=8.8cm]{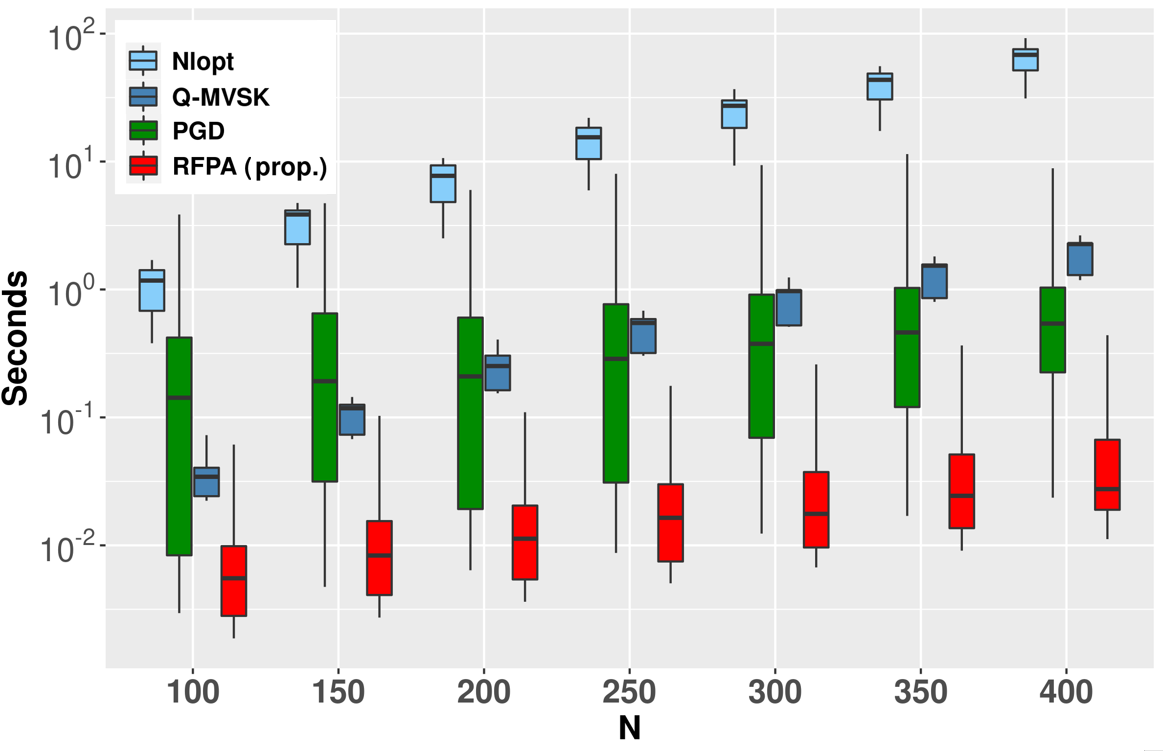}
\par\end{centering}
\caption{Comparison of algorithms with respect to the computational time under
different data dimension. \label{fig:Comparison-of-scalability-onMVSK}}
\end{figure}

According to the numerical simulation results shown in Figure \ref{fig:Comparison-of-scalability-onMVSK},
our proposed outperforms the state-of-the-art methods by one-to-two
orders of magnitude when we assume the data follows a ghMST distribution.
The difference seems to be enlarged when the problem dimension increases.
Besides, the RFPA algorithm appears to be more stable compared to
the PGD method.

\subsubsection{Comparison on scalability}

Interestingly, implied by Figure \ref{fig:Comparison-of-scalability-onMVSK},
first-order methods, including RFPA and PGD, appear to be more scalable
than the second-order Q-MVSK algorithm. To better investigate this
phenomenon, we will be conducting a comparison of these algorithms
using a synthetic data set, where the parameter $\boldsymbol{\Theta}$
is randomly generated. 

\begin{table}
\caption{Empirical orders of complexity. \label{tab:Empirical-orders-of}}

\centering{}%
\begin{tabular}{|c|c|c|c|}
\hline 
Q-MVSK & Nlopt & PGD & RFPA\tabularnewline
\hline 
2.864 & 3.827 & 1.976 & 1.944\tabularnewline
\hline 
\end{tabular}
\end{table}

\begin{figure}
\begin{centering}
\includegraphics[width=8.8cm]{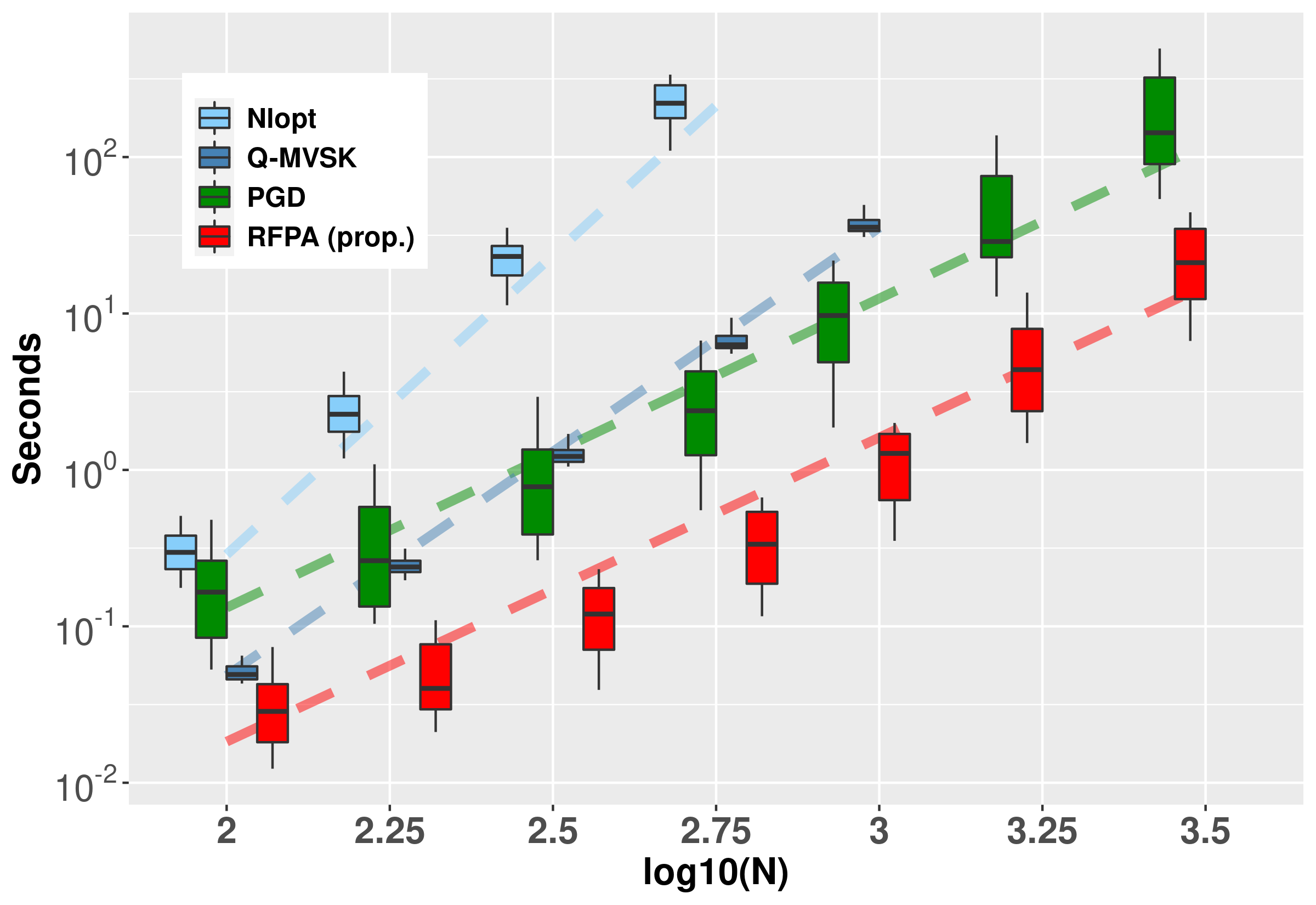}
\par\end{centering}
\caption{Investigation on the empirical complexity of RFPA and Q-MVSK algorithms.
\label{fig:complexity}}
\end{figure}

As shown in Figure \ref{fig:complexity}, the proposed RFPA algorithm
has a significantly lower complexity compared to the Q-MVSK algorithm,
as its every single iteration does not contain procedures with high
complexity. Meanwhile, PGD method also enjoys the benefits of low
complexity but its overall efficiency is worse than the RFPA method.
We also fit the empirical orders of the four methods considered. The
relative results are shown in Table \ref{tab:Empirical-orders-of}.
It turns out that the empirical computational complexity of our method
is $\mathcal{O}\left(N^{2}\right)$ and the complexity of the second-order
method Q-MVSK is around $\mathcal{O}\left(N^{3}\right)$. The results
of numerical simulations coincide with the discussion in Section \ref{subsec:Complexity-Analysis}. 

\subsection{Empirical Convergence of the proposed RFPA algorithm \label{subsec:Empirical-Convergence}}

\begin{figure}
\begin{centering}
\includegraphics[width=8.8cm]{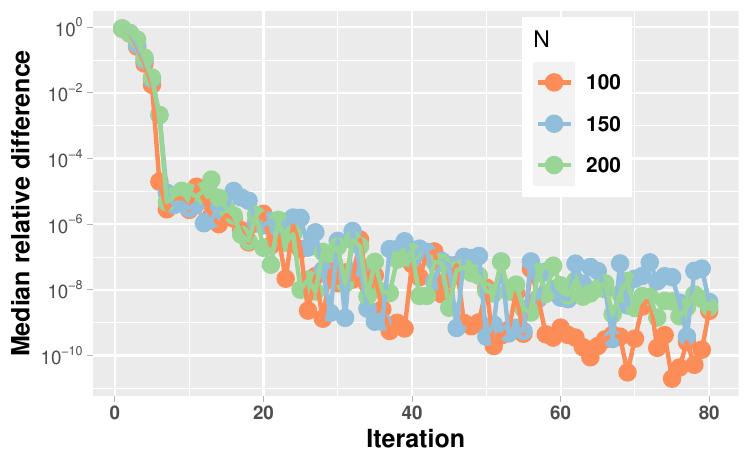}
\par\end{centering}
\caption{Median relative difference at each iteration. \label{fig:RD}}
\end{figure}

According to Theorem \ref{thm:If-,-then}, when $\mathbf{w}^{k}=\mathbf{w}^{k+1}$,
the algorithm terminates at a stationary point of Problem (\ref{eq:MVSK}).
Though exact equality is often unattainable, empirically, the relative
difference of $\mathbf{w}$, denoted as 
\begin{equation}
\text{Relative Error}\left(\mathbf{w}^{k}\right)\overset{\Delta}{=}\left\Vert \mathbf{w}^{k}-\mathbf{w}^{k-1}\right\Vert \big/\left\Vert \mathbf{w}^{k}\right\Vert ,\label{eq:ddd}
\end{equation}
would tend to zero. To show this, we conduct experiments using real-world
data sets with different problem dimensions. The values of (\ref{eq:ddd})
are computed at each iteration. From \ref{fig:RD} we observe that
the differences all reduce to very small numbers. Empirical studies
show that the residual value $R\left(\mathbf{w}^{k};\eta\right)$
would tend to zero after $20$ iterations and the final solution would
converge to the stationary point of Problem (\ref{eq:MVSK}).

\section{Conclusion \label{sec:Conclusion}}

In this paper, we have proposed a high-order portfolio design framework
with the help of the parametric skew-$t$ distribution and a robust
fixed point acceleration. The parametric approach is practical for
modeling the skewness and kurtosis of portfolio returns in high-dimensional
settings. By assuming the returns follow a ghMST distribution, we
can alleviate the difficulties caused by the high complexity of traditional
methods and accelerate all existing algorithms to a certain extent.
Additionally, the proposed RFPA algorithm immensely cut down the number
of iterations for first-order methods. Numerical simulations have
demonstrated the outstanding efficiency and scalability of our proposed
framework over the state-of-the-are benchmarks. 

\appendix{}

\subsection{Computational time of different estimation methods \label{subsec:Comparing_estimation_time}}

\begin{figure}[hbt!]
\begin{center}
\includegraphics[width= 8cm]{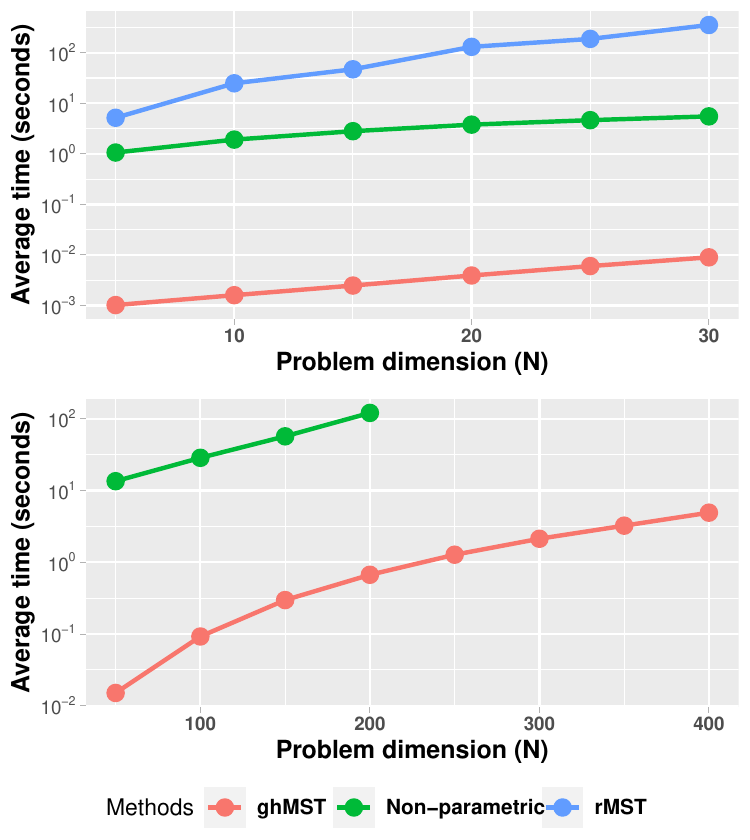}
\caption{Comparing computational time (seconds) of different estimation methods.}
\label{fig:estimation_time}
\end{center}
\end{figure}Figure \ref{fig:estimation_time} depicts the computational time of
different estimation methods. It can be observed that fitting the
ghMST distribution is much more efficient than others. 

\subsection{Proof for Lemma \ref{lem:Skew_t_matrix} \label{subsec:Proof-for-Lemma 1}}

The proof starts with a fact that the central moments of a Gaussian
variable $\tilde{\mathbf{X}}\sim\mathcal{N}\left(\tilde{\boldsymbol{\mu}},\tilde{\boldsymbol{\Sigma}}\right)$
is given by 
\begin{equation}
\begin{aligned}\mathbb{E}[\tilde{X}_{i}] & =\text{ }\tilde{\mu}_{i},\\
\mathbb{E}[\tilde{X}_{i}\tilde{X}_{j}] & =\text{ }\tilde{\mu}_{i}\tilde{\mu}_{j}+\tilde{\Sigma}_{ij},\\
\mathbb{E}[\tilde{X}_{i}\tilde{X}_{j}\tilde{X}_{k}] & =\text{ }\tilde{\mu}_{i}\tilde{\mu}_{j}\tilde{\mu}_{k}+\tilde{\mu}_{i}\tilde{\Sigma}_{jk}+\tilde{\mu}_{j}\tilde{\Sigma}_{ik}+\tilde{\mu}_{k}\tilde{\Sigma}_{ij},\\
\mathbb{E}[\tilde{X}_{i}\tilde{X}_{j}\tilde{X}_{k}\tilde{X}_{l}] & =\text{ }\tilde{\mu}_{i}\tilde{\mu}_{j}\tilde{\mu}_{k}\tilde{\mu}_{l}+\underbrace{(\tilde{\Sigma}_{ij}\tilde{\mu}_{k}\tilde{\mu}_{l}+\cdots+\tilde{\Sigma}_{kl}\tilde{\mu}_{i}\tilde{\mu}_{j})}_{6\text{ items}}\\
 & \quad+(\tilde{\Sigma}_{ij}\tilde{\Sigma}_{kl}+\tilde{\Sigma}_{ik}\tilde{\Sigma}_{jl}+\tilde{\Sigma}_{il}\tilde{\Sigma}_{jk}).
\end{aligned}
\label{eq:s}
\end{equation}
 Then, given the first term of the hierarchical structure $\mathbf{r}|\tau\overset{\text{i.i.d}}{\sim}\mathcal{N}\left(\boldsymbol{\mu}+\frac{1}{\tau}\boldsymbol{\gamma},\frac{1}{\tau}\boldsymbol{\Sigma}\right)$,
we have 
\begin{equation}
\mathbb{E}\left[r_{i}|\tau\right]=\mu_{i}+\frac{1}{\tau}\gamma_{i},\label{eq:Eri}
\end{equation}
\begin{equation}
\mathbb{E}\left[r_{i}\right]=\mu_{i}+\mathbb{E}\left[\frac{1}{\tau}\right]\gamma_{i}=\mu_{i}+\frac{\nu}{\nu-2}\gamma_{i}.\label{eq:Eri2}
\end{equation}
Meanwhile, the hierarchical structure can be further written as 
\begin{equation}
\begin{aligned}\mathbf{r}|\tau-\mathbb{E}\left[\mathbf{r}|\tau\right]\,\, & \overset{\text{i.i.d}}{\sim}\,\,\mathcal{N}\left(\boldsymbol{\mu}+\frac{1}{\tau}\boldsymbol{\gamma}-\mathbb{E}\left[\mathbf{r}|\tau\right],\frac{1}{\tau}\boldsymbol{\Sigma}\right),\\
\tau\,\, & \overset{\text{i.i.d}}{\sim}\,\,\text{Gamma}\left(\frac{\nu}{2},\frac{\nu}{2}\right),
\end{aligned}
\label{eq:rtau}
\end{equation}
where $\boldsymbol{\mu}+\frac{1}{\tau}\boldsymbol{\gamma}-\mathbb{E}\left[\mathbf{r}|\tau\right]=\left(\frac{1}{\tau}-\frac{\nu}{\nu-2}\right)\boldsymbol{\gamma}.$
Therefore, we can compute the central moments of $\tilde{\mathbf{r}}|\tau=\mathbf{r}|\tau-\mathbb{E}\left[\mathbf{r}|\tau\right]$
by regarding $\tilde{\boldsymbol{\mu}}=\left(\frac{1}{\tau}-\frac{\nu}{\nu-2}\right)\boldsymbol{\gamma}$
and $\tilde{\boldsymbol{\Sigma}}=\frac{1}{\tau}\boldsymbol{\Sigma}$:
\begin{align}
\mathbb{E}[\left.\tilde{r}_{i}\tilde{r}_{j}\right|\tau] & =\left(\frac{1}{\tau}-\frac{\nu}{\nu-2}\right)^{2}\gamma_{i}\gamma_{j}+\frac{1}{\tau}\Sigma_{ij},\nonumber \\
\mathbb{E}[\left.\tilde{r}_{i}\tilde{r}_{j}\tilde{r}_{k}\right|\tau] & =\left(\frac{1}{\tau}-\frac{\nu}{\nu-2}\right)^{3}\gamma_{i}\gamma_{j}\gamma_{k}+\frac{1}{\tau}\left(\frac{1}{\tau}-\frac{\nu}{\nu-2}\right)\cdot\nonumber \\
 & \quad\quad\left[\gamma_{i}\Sigma_{jk}+\gamma_{j}\Sigma_{ik}+\gamma_{k}\Sigma_{ij}\right],\nonumber \\
\mathbb{E}[\left.\tilde{r}_{i}\tilde{r}_{j}\tilde{r}_{k}\tilde{r}_{l}\right|\tau] & =\left(\frac{1}{\tau}-\frac{\nu}{\nu-2}\right)^{4}\gamma_{i}\gamma_{j}\gamma_{k}\gamma_{l}+\nonumber \\
 & \quad\left(\frac{1}{\tau}-\frac{\nu}{\nu-2}\right)^{2}\frac{1}{\tau}\cdot\nonumber \\
 & \quad\underbrace{(\Sigma_{ij}\gamma_{k}\gamma_{l}+\cdots+\Sigma_{kl}\gamma_{i}\gamma_{j})}_{6\text{ items}}+\nonumber \\
 & \quad\frac{1}{\tau^{2}}(\Sigma_{ij}\Sigma_{kl}+\Sigma_{ik}\Sigma_{jl}+\Sigma_{il}\Sigma_{jk}).\label{eq:Errr}
\end{align}
By taking expectation subject to $\tau$, i.e, $\mathbb{E}\left[\tau^{-1}\right]=\frac{\nu}{\nu-2}$,
$\mathbb{E}\left[\tau^{-2}\right]=\frac{\nu^{2}}{\left(\nu-2\right)\left(\nu-4\right)},$$\mathbb{E}\left[\tau^{-3}\right]=\frac{\nu^{3}}{\left(\nu-2\right)\left(\nu-4\right)\left(\nu-6\right)}$,
and $\mathbb{E}\left[\tau^{-4}\right]=\frac{\nu^{4}}{\left(\nu-2\right)\left(\nu-4\right)\left(\nu-6\right)\left(\nu-8\right)}$,
the Lemma \ref{lem:Skew_t_matrix} is obtained. 

\subsection{Proof for Lemma \ref{lem:portfolio_high_order_moments} \label{subsec:Proof-for-Lemma 2}}

Assuming $\mathbf{r}\sim\textsf{ghMST}\left(\boldsymbol{\mu},\boldsymbol{\Sigma},\boldsymbol{\gamma},\nu\right)$,
which indicates that the portfolio return $\mathbf{w}^{T}\mathbf{r}$
satisfies the following hierarchical structure:
\begin{equation}
\begin{aligned}\mathbf{w}^{T}\mathbf{r}|\tau\,\, & \overset{\text{i.i.d}}{\sim}\,\,\mathcal{N}\left(\mathbf{w}^{T}\boldsymbol{\mu}+\frac{1}{\tau}\mathbf{w}^{T}\boldsymbol{\gamma},\frac{1}{\tau}\mathbf{w}^{T}\boldsymbol{\Sigma}\mathbf{w}\right),\\
\tau\,\, & \overset{\text{i.i.d}}{\sim}\,\,\text{Gamma}\left(\frac{\nu}{2},\frac{\nu}{2}\right),
\end{aligned}
\label{eq:structure_of_portfolio}
\end{equation}
Then, according to (\ref{eq:structure_of_portfolio}), we have 
\begin{equation}
\mathbf{w}^{T}\mathbf{r}\sim\text{\textsf{ghMST}\ensuremath{\left(\mathbf{w}^{T}\boldsymbol{\mu},\mathbf{w}^{T}\boldsymbol{\Sigma}\mathbf{w},\mathbf{w}^{T}\boldsymbol{\gamma},\nu\right)}}.\label{eq:wtr}
\end{equation}
As $\mathbf{w}^{T}\mathbf{r}$ is a scalar, its high-order central
moments, i.e., $\boldsymbol{\Phi}$ and $\boldsymbol{\Psi}$ are all
scalars. Based on Lemma \ref{lem:Skew_t_matrix}, we replace $\boldsymbol{\mu}$,
$\boldsymbol{\Sigma}$, and $\boldsymbol{\gamma}$ with $\mathbf{w}^{T}\boldsymbol{\mu},\mathbf{w}^{T}\boldsymbol{\Sigma}\mathbf{w}$,
and $\mathbf{w}^{T}\boldsymbol{\gamma}$, respectively. Then we can
obtain 
\begin{align}
\phi_{3}\left(\mathbf{w}\right) & =\Phi=\mathbb{E}\left[\left(\frac{1}{\tau}-\frac{\nu}{\nu-2}\right)^{3}\right]\left(\mathbf{w}^{T}\boldsymbol{\gamma}\right)^{3}+\nonumber \\
 & \quad\quad3\mathbb{E}\left[\frac{1}{\tau}\left(\frac{1}{\tau}-\frac{\nu}{\nu-2}\right)\right]\left(\mathbf{w}^{T}\boldsymbol{\Sigma}\mathbf{w}\cdot\mathbf{w}^{T}\boldsymbol{\gamma}\right),\nonumber \\
\phi_{4}\left(\mathbf{w}\right) & =\Psi=\mathbb{E}\left[\left(\frac{1}{\tau}-\frac{\nu}{\nu-2}\right)^{4}\right]\left(\mathbf{w}^{T}\boldsymbol{\gamma}\right)^{4}\nonumber \\
 & \quad\quad+6\mathbb{E}\left[\left(\frac{1}{\tau}-\frac{\nu}{\nu-2}\right)^{2}\right]\left(\mathbf{w}^{T}\boldsymbol{\gamma}\right)^{2}\left(\mathbf{w}^{T}\boldsymbol{\Sigma}\mathbf{w}\right)\nonumber \\
 & \quad\quad+3\mathbb{E}\left[\frac{1}{\tau^{2}}\right]a_{43}\left(\mathbf{w}^{T}\boldsymbol{\Sigma}\mathbf{w}\right)^{2}.\label{eq:phi34(w)}
\end{align}
Simply follows the definition of $\mathbf{a}$, Lemma \ref{lem:portfolio_high_order_moments}
is proved. 

\subsection{Gradient and Hessian of the high-order moments \label{subsec:Gradient-and-Hessian}}

Based on Lemma \ref{lem:portfolio_high_order_moments}, the gradient
and Hessian of the skewness and kurtosis subject to $\mathbf{w}$
can be computed as
\begin{equation}
\begin{array}{rl}
\nabla\phi_{3}\left(\mathbf{w}\right)= & 3a_{31}\left(\mathbf{w}^{T}\boldsymbol{\gamma}\right)^{2}\boldsymbol{\gamma}\\
 & +a_{32}\left[\left(\mathbf{w}^{T}\boldsymbol{\Sigma}\mathbf{w}\right)\boldsymbol{\gamma}+2\left(\mathbf{w}^{T}\boldsymbol{\gamma}\right)\boldsymbol{\Sigma}\mathbf{w}\right],\\
\nabla^{2}\phi_{3}\left(\mathbf{w}\right)= & 6a_{31}\left(\mathbf{w}^{T}\boldsymbol{\gamma}\right)\boldsymbol{\gamma}\boldsymbol{\gamma}^{T}\\
 & +2a_{32}\left[\boldsymbol{\gamma}\mathbf{w}^{T}\boldsymbol{\Sigma}+\boldsymbol{\Sigma}\mathbf{w}\boldsymbol{\gamma}^{T}+\mathbf{w}^{T}\boldsymbol{\gamma}\boldsymbol{\Sigma}\right],\\
\nabla\phi_{4}\left(\mathbf{w}\right)= & 4a_{41}\left(\mathbf{w}^{T}\boldsymbol{\gamma}\right)^{3}\boldsymbol{\gamma}\\
 & +2a_{42}\left[\left(\mathbf{w}^{T}\boldsymbol{\gamma}\right)^{2}\boldsymbol{\Sigma}\mathbf{w}+\left(\mathbf{w}^{T}\boldsymbol{\Sigma}\mathbf{w}\right)\left(\mathbf{w}^{T}\boldsymbol{\gamma}\right)\boldsymbol{\gamma}\right]\\
 & +4a_{43}\left(\mathbf{w}^{T}\boldsymbol{\Sigma}\mathbf{w}\right)\boldsymbol{\Sigma}\mathbf{w},\\
\nabla^{2}\phi_{4}\left(\mathbf{w}\right)= & 12a_{41}\left(\mathbf{w}^{T}\boldsymbol{\gamma}\right)^{2}\boldsymbol{\gamma}\boldsymbol{\gamma}^{T}\\
 & +2a_{42}\left[2\left(\mathbf{w}^{T}\boldsymbol{\gamma}\right)\boldsymbol{\Sigma}\mathbf{w}\boldsymbol{\gamma}^{T}+\left(\mathbf{w}^{T}\boldsymbol{\gamma}\right)^{2}\boldsymbol{\Sigma}+\right.\\
 & \quad\quad\left.2\left(\mathbf{w}^{T}\boldsymbol{\gamma}\right)\boldsymbol{\gamma}\mathbf{w}^{T}\boldsymbol{\Sigma}+\left(\mathbf{w}^{T}\boldsymbol{\Sigma}\mathbf{w}\right)\boldsymbol{\gamma}\boldsymbol{\gamma}^{T}\right]\\
 & +4a_{43}\left[2\boldsymbol{\Sigma}\mathbf{w}\mathbf{w}^{T}\boldsymbol{\Sigma}+\left(\mathbf{w}^{T}\boldsymbol{\Sigma}\mathbf{w}\right)\boldsymbol{\Sigma}\right].
\end{array}\label{eq:gandH}
\end{equation}

\subsection{Water-filling algorithm \label{subsec:Waterfilling-algorithm}}

Here we consider an optimization problem
\begin{equation}
\begin{array}{ll}
\underset{\mathbf{w}}{\mathsf{minimize}}\, & \frac{1}{2}\left\Vert \mathbf{w}-\left(\mathbf{w}^{k}-\eta\nabla f\left(\mathbf{w}^{k}\right)\right)\right\Vert _{2}^{2}\\
\mathsf{subject}\text{ }\mathsf{to} & \mathbf{w}\in\mathcal{W}.
\end{array}\label{eq:projection_problem}
\end{equation}
Given $\mathcal{W}=\left\{ \mathbf{w}\left|\mathbf{1}^{T}\mathbf{w}=1,\mathbf{w}\geq\mathbf{0}\right.\right\} $,
the Lagrangian of Problem (\ref{eq:projection_problem}) is 
\begin{align}
\mathcal{L}\left(\mathbf{w},\boldsymbol{\psi},\gamma\right) & =\frac{1}{2}\left\Vert \mathbf{w}-\left(\mathbf{w}^{k}-\eta\nabla f\left(\mathbf{w}^{k}\right)\right)\right\Vert _{2}^{2}\nonumber \\
 & \quad-\boldsymbol{\psi}^{T}\mathbf{w}+\gamma\left(\mathbf{1}^{T}\mathbf{w}-1\right),\label{eq:L}
\end{align}
where $\boldsymbol{\psi}$ and $\gamma$ are dual variables associated
with the constraints $\mathbf{w}\geq\mathbf{0}$ and $\mathbf{1}^{T}\mathbf{w}=1$,
respectively. The KKT conditions are
\begin{align}
\eta\nabla f\left(\mathbf{w}^{k}\right)+\left(\mathbf{w}-\mathbf{w}^{k}\right)-\boldsymbol{\psi}+\gamma\mathbf{1} & =\mathbf{0},\nonumber \\
\boldsymbol{\psi}\odot\mathbf{w} & =\mathbf{0}.\label{eq:kkt}
\end{align}
Hence, we have 
\begin{equation}
w_{i}=\max\left(0,w_{i}^{k}-\eta\left[\nabla f\left(\mathbf{w}^{k}\right)\right]_{i}-\gamma\right).\label{eq:wi}
\end{equation}
Define a continuous and monotone decreasing function $\zeta:\mathbb{R}\rightarrow\mathbb{R}$:
\begin{equation}
\zeta\left(\gamma\right)=\sum_{i=1}^{N}\max\left(0,w_{i}^{k}-\eta\left[\nabla f\left(\mathbf{w}^{k}\right)\right]_{i}-\gamma\right)-1\label{eq:xigamma}
\end{equation}
with $\zeta\left(-\infty\right)=+\infty$ and $\zeta\left(-\infty\right)=-1$,
the root 
\begin{equation}
\gamma^{\star}=\arg\left(\zeta\left(\gamma\right)=0\right)\label{eq:gammastar}
\end{equation}
exists and is unique. The root provides a dual optimal of the KKT
system. We can easily solve $\gamma$ and $\mathbf{w}^{\star}$ via
bisection. 

\subsection{Monotonicity of the sequence $\left\{ f\left(\mathbf{w}^{k}\right)\right\} $
\label{subsec:Monotonicity-of-the}}

According to the projection theorem \cite{facchinei2003finite}, i.e.,
\begin{equation}
\forall\mathbf{x},\mathbf{z}:\left\langle \mathbf{z}-\mathbf{x},\mathcal{P}_{\mathcal{W}}\left(\mathbf{z}\right)-\mathcal{P}_{\mathcal{W}}\left(\mathbf{x}\right)\right\rangle \geq\left\Vert \mathcal{P}_{\mathcal{W}}\left(\mathbf{z}\right)-\mathcal{P}_{\mathcal{W}}\left(\mathbf{x}\right)\right\Vert _{2}^{2},\label{eq:projection_thm}
\end{equation}
we apply $\mathbf{z}=\mathbf{w}^{k}-\eta\nabla f\left(\mathbf{w}^{k}\right)$
and $\mathbf{x}=\mathbf{w}^{k}$ to obtain
\begin{equation}
\left\langle -\eta\nabla f\left(\mathbf{w}^{k}\right),\mathbf{w}^{k+1}-\mathbf{w}^{k}\right\rangle \geq\left\Vert \mathbf{w}^{k+1}-\mathbf{w}^{k}\right\Vert _{2}^{2},\label{eq:pjtm}
\end{equation}
or equivalently
\begin{equation}
\left\langle \nabla f\left(\mathbf{w}^{k}\right),\mathbf{w}^{k+1}-\mathbf{w}^{k}\right\rangle \leq-\frac{1}{\eta}\left\Vert \mathbf{w}^{k+1}-\mathbf{w}^{k}\right\Vert _{2}^{2}.\label{eq:ssss}
\end{equation}
Hence, from the inequality (\ref{eq:line_search}) we have
\begin{align}
f\left(\mathbf{w}^{k+1}\right) & \leq f\left(\mathbf{w}^{k}\right)+\nabla f\left(\mathbf{w}^{k}\right)^{T}\left(\mathbf{w}^{k+1}-\mathbf{w}^{k}\right)\nonumber \\
 & \quad+\frac{1}{2\eta}\left\Vert \mathbf{w}^{k}-\mathbf{w}^{k+1}\right\Vert _{2}^{2}\nonumber \\
 & \leq f\left(\mathbf{w}^{k}\right)-\frac{1}{\eta}\left\Vert \mathbf{w}^{k+1}-\mathbf{w}^{k}\right\Vert _{2}^{2}\nonumber \\
 & \quad+\frac{1}{2\eta}\left\Vert \mathbf{w}^{k}-\mathbf{w}^{k+1}\right\Vert _{2}^{2}\nonumber \\
 & =f\left(\mathbf{w}^{k}\right)-\frac{1}{2\eta}\left\Vert \mathbf{w}^{k}-\mathbf{w}^{k+1}\right\Vert _{2}^{2}\leq f\left(\mathbf{w}^{k}\right),\label{eq:mono1}
\end{align}
 which indicates that the sequence $\left\{ f\left(\mathbf{w}^{k}\right)\right\} $
is then monotone.

\subsection{Proof of Theorem \ref{thm:If-,-then} \label{subsec:Proof-of-Theorem 1}}
\begin{IEEEproof}
When $\mathbf{w}^{k}=\mathbf{w}^{k+1}$, we may have $\mathbf{w}^{k+1}=\mathcal{P}_{\mathcal{W}}\left(\mathbf{w}^{k}-2\alpha^{k}R\left(\mathbf{w}^{k};\eta\right)+\left(\alpha^{k}\right)^{2}V\left(\mathbf{w}^{k};\eta\right)\right)$
or $\mathbf{w}^{k+1}=G\left(\mathbf{w}^{k};\eta'\right)$.

(i) We first analyze the first case where $\mathbf{w}^{k+1}=\mathcal{P}_{\mathcal{W}}\left(\mathbf{y}^{k}\right)$,
in which 
\begin{equation}
\mathbf{y}^{k}\overset{\Delta}{=}\mathbf{w}^{k}-2\alpha^{k}R\left(\mathbf{w}^{k};\eta\right)+\left(\alpha^{k}\right)^{2}V\left(\mathbf{w}^{k};\eta\right).\label{eq:ykd}
\end{equation}
By applying the contraposition, we prove the following statement instead
\begin{equation}
\forall\mathbf{w}^{k}\in\mathcal{W}:R\left(\mathbf{w}^{k};\eta\right)\neq\mathbf{0}\Rightarrow\mathcal{P}_{\mathcal{W}}\left(\mathbf{y}^{k}\right)\neq\mathbf{w}^{k}.\label{eq:wkR}
\end{equation}
For simplicity, we denote $\alpha=-\alpha^{k}>0$. Note that $\alpha\neq0$
as $R\left(\mathbf{w}^{k};\eta\right)\neq\mathbf{0}$.

(A) If $\alpha\in\left(0,1\right]$, then, we obtain
\begin{align}
\mathbf{y}^{k} & =\left(1-2\alpha+\alpha^{2}\right)\mathbf{w}^{k}+\left(2\alpha-2\alpha^{2}\right)G\left(\mathbf{w}^{k};\eta\right)\nonumber \\
 & \quad+\alpha^{2}G\left(G\left(\mathbf{w}^{k};\eta\right);\eta\right)\nonumber \\
 & \overset{\Delta}{=}a^{k}\mathbf{w}^{k}+b^{k}G\left(\mathbf{w}^{k};\eta\right)+c^{k}G\left(G\left(\mathbf{w}^{k};\eta\right);\eta\right)\label{eq:yk123}
\end{align}
in which $a^{k}=1-2\alpha+\alpha^{2}$, $b^{k}=2\alpha-2\alpha^{2}$,
and $c^{k}=\alpha^{2}$. As $0<\alpha\leq1$, we have $0\leq a^{k}<1$,
$0\leq b^{k}\leq\frac{1}{2}$, $0<c^{k}\leq1$, and $a^{k}+b^{k}+c^{k}=1$.
Hence, $\mathbf{y}^{k}$ is a convex combination of $\mathbf{w}^{k}$,
$G\left(\mathbf{w}^{k};\eta\right)$, and $G\left(G\left(\mathbf{w}^{k};\eta\right);\eta\right)$.
As a result, $\mathbf{y}^{k}\in\mathcal{W}$ and the projection of
$\mathbf{y}^{k}$ onto $\mathcal{W}$ is itself, i.e., $\mathcal{P}_{\mathcal{W}}\left(\mathbf{y}^{k}\right)=\mathbf{y}^{k}$.
Consequently, we obtain
\begin{equation}
\mathbf{w}^{k+1}=\mathcal{P}_{\mathcal{W}}\left(\mathbf{y}^{k}\right)\neq\mathbf{w}^{k}.\label{eq:ow}
\end{equation}

(B) If $\alpha\in\left(1,\infty\right)$. We will first show that
the following inequality holds for any $\mathbf{w}^{k}$ 
\begin{equation}
\xi\overset{\Delta}{=}\left\langle R\left(\mathbf{w}^{k};\eta\right),R\left(\mathbf{w}^{k};\eta\right)+\alpha V\left(\mathbf{w}^{k};\eta\right)\right\rangle \geq0.\label{eq:critical_inequality}
\end{equation}
In principle, we consider the following three cases based on the value
of $\left\langle R\left(\mathbf{w}^{k};\eta\right),V\left(\mathbf{w}^{k};\eta\right)\right\rangle $.

(B.1) If $\left\langle R\left(\mathbf{w}^{k};\eta\right),V\left(\mathbf{w}^{k};\eta\right)\right\rangle \geq0$,
then $b\left(\mathbf{w}^{k}\right)=-\infty$. (\ref{eq:critical_inequality})
holds as $\forall\alpha>1$:
\begin{equation}
\xi=\left\Vert R\left(\mathbf{w}^{k};\eta\right)\right\Vert ^{2}+\alpha\left\langle R\left(\mathbf{w}^{k};\eta\right),V\left(\mathbf{w}^{k};\eta\right)\right\rangle \ge0.\label{eq:xx}
\end{equation}

(B.2) If $\left\langle R\left(\mathbf{w}^{k};\eta\right),V\left(\mathbf{w}^{k};\eta\right)\right\rangle <0$
and $b\left(\mathbf{w}^{k}\right)=\frac{\left\Vert R\left(\mathbf{w}^{k};\eta\right)\right\Vert _{2}^{2}}{\left\langle R\left(\mathbf{w}^{k};\eta\right),V\left(\mathbf{w}^{k};\eta\right)\right\rangle }\neq-\infty$,
we have $\alpha\leq-b\left(\mathbf{w}^{k}\right)$. In this case,
(\ref{eq:critical_inequality}) holds as $\forall\alpha\in\left(1,-b\left(\mathbf{w}^{k}\right)\right)$:
\begin{align}
\xi= & \left\Vert R\left(\mathbf{w}^{k};\eta\right)\right\Vert ^{2}+\alpha\left\langle R\left(\mathbf{w}^{k};\eta\right),V\left(\mathbf{w}^{k};\eta\right)\right\rangle \nonumber \\
\geq & \left\Vert R\left(\mathbf{w}^{k};\eta\right)\right\Vert ^{2}-b\left(\mathbf{w}^{k}\right)\left\langle R\left(\mathbf{w}^{k};\eta\right),V\left(\mathbf{w}^{k};\eta\right)\right\rangle =0.\label{eq:xx1}
\end{align}

(B.3) If $\left\langle R\left(\mathbf{w}^{k};\eta\right),V\left(\mathbf{w}^{k};\eta\right)\right\rangle <0$
but $b\left(\mathbf{w}^{k}\right)\rightarrow-\infty$ due to $V\left(\mathbf{w}^{k};\eta\right)\rightarrow\mathbf{0}$,
the value of $\alpha$ can be either $\left\Vert R\left(\mathbf{w}^{k};\eta\right)\right\Vert \big/\left\Vert V\left(\mathbf{w}^{k};\eta\right)\right\Vert \rightarrow\infty$
or $-b\left(\mathbf{w}^{k}\right)\rightarrow\infty$. When $\alpha=\left\Vert R\left(\mathbf{w}^{k};\eta\right)\right\Vert \big/\left\Vert V\left(\mathbf{w}^{k};\eta\right)\right\Vert $,
we suppose 
\begin{align}
 & \left\langle R\left(\mathbf{w}^{k};\eta\right),V\left(\mathbf{w}^{k};\eta\right)\right\rangle \nonumber \\
= & \left\Vert R\left(\mathbf{w}^{k};\eta\right)\right\Vert \left\Vert V\left(\mathbf{w}^{k};\eta\right)\right\Vert \cos\theta_{R,V},\label{eq:xx2}
\end{align}
in which $\theta_{R,V}$ is the angle between $R\left(\mathbf{w}^{k};\eta\right)$
and $V\left(\mathbf{w}^{k};\eta\right)$. Hence, as $\cos\theta_{R,V}\in\left[-1,1\right]$,
we obtain
\begin{align}
\xi & =\left\Vert R\left(\mathbf{w}^{k};\eta\right)\right\Vert ^{2}+\frac{\left\Vert R\left(\mathbf{w}^{k};\eta\right)\right\Vert }{\left\Vert V\left(\mathbf{w}^{k};\eta\right)\right\Vert }\left\langle R\left(\mathbf{w}^{k};\eta\right),V\left(\mathbf{w}^{k};\eta\right)\right\rangle \nonumber \\
 & =\left\Vert R\left(\mathbf{w}^{k};\eta\right)\right\Vert ^{2}\left(1+\cos\theta_{R,V}\right)\geq0.\label{eq:xx3}
\end{align}
When $\alpha=-b\left(\mathbf{w}^{k}\right)=-\frac{\left\Vert R\left(\mathbf{w}^{k};\eta\right)\right\Vert _{2}^{2}}{\left\langle R\left(\mathbf{w}^{k};\eta\right),V\left(\mathbf{w}^{k};\eta\right)\right\rangle }$,
it is obvious that 
\begin{equation}
\xi=\left\Vert R\left(\mathbf{w}^{k};\eta\right)\right\Vert ^{2}-b\left(\mathbf{w}^{k}\right)\left\langle R\left(\mathbf{w}^{k};\eta\right),V\left(\mathbf{w}^{k};\eta\right)\right\rangle =0.\label{eq:xx4}
\end{equation}

Therefore, (\ref{eq:critical_inequality}) holds. As a consequence,
we can compare the following two terms
\begin{align}
\mathbf{y}^{k}-\mathbf{w}^{k} & =\alpha R\left(\mathbf{w}^{k};\eta\right)\nonumber \\
 & \quad+\alpha\left(R\left(\mathbf{w}^{k};\eta\right)+\alpha V\left(\mathbf{w}^{k};\eta\right)\right),\nonumber \\
\mathbf{y}^{k}-G\left(\mathbf{w}^{k};\eta\right) & =\left(\alpha-1\right)R\left(\mathbf{w}^{k};\eta\right)\nonumber \\
 & \quad+\alpha\left(R\left(\mathbf{w}^{k};\eta\right)+\alpha V\left(\mathbf{w}^{k};\eta\right)\right),\label{eq:ydis}
\end{align}
by evaluating the difference of their squared $\ell_{2}$ norms, i.e.,
\begin{align}
 & \left\Vert \mathbf{y}^{k}-\mathbf{w}^{k}\right\Vert ^{2}-\left\Vert \mathbf{y}^{k}-G\left(\mathbf{w}^{k};\eta\right)\right\Vert ^{2}\nonumber \\
= & \left(2\alpha-1\right)\left\Vert R\left(\mathbf{w}^{k};\eta\right)\right\Vert ^{2}\nonumber \\
 & \quad+2\alpha\left\langle R\left(\mathbf{w}^{k};\eta\right),R\left(\mathbf{w}^{k};\eta\right)+\alpha V\left(\mathbf{w}^{k};\eta\right)\right\rangle .\label{eq:yds}
\end{align}
Then, we obtain the following strict inequality
\begin{equation}
\left\Vert \mathbf{y}^{k}-\mathbf{w}^{k}\right\Vert ^{2}-\left\Vert \mathbf{y}^{k}-G\left(\mathbf{w}^{k};\eta\right)\right\Vert ^{2}>0\label{eq:tc}
\end{equation}
as $\alpha>1$ and $\left\Vert R\left(\mathbf{w}^{k};\eta\right)\right\Vert >0$.
Therefore, $\mathcal{P}_{\mathcal{W}}\left(\mathbf{y}^{k}\right)\neq\mathbf{w}^{k}$
as there exists a feasible point $G\left(\mathbf{w}^{k};\eta\right)\in\mathcal{W}$
that is closer to $\mathbf{y}^{k}$ compared to $\mathbf{w}^{k}$. 

Hence, we have shown that $\mathcal{P}_{\mathcal{W}}\left(\mathbf{y}^{k}\right)\neq\mathbf{w}^{k}$
if $R\left(\mathbf{w}^{k};\eta\right)\neq\mathbf{0}$. As a result,
we have obtained the following statement
\begin{equation}
\mathcal{P}_{\mathcal{W}}\left(\mathbf{y}^{k}\right)=\mathbf{w}^{k}\Rightarrow R\left(\mathbf{w}^{k};\eta\right)=\mathbf{0}.\label{eq:ob}
\end{equation}
Then, $\mathbf{w}^{k}$ is a stationary point of Problem (\ref{eq:MVSK})
according to Lemma \ref{lem:The-set-of-fixed-point}.

(ii) We then analyze the second case where 
\begin{equation}
\mathbf{w}^{k+1}=\mathcal{P}_{\mathcal{W}}\left(\mathbf{w}^{k}-\eta'\nabla f\left(\mathbf{w}^{k}\right)\right).\label{eq:yp}
\end{equation}
Then, $\mathbf{w}^{k}$ is a stationary point of Problem (\ref{eq:MVSK})
when $\mathbf{w}^{k+1}=\mathbf{w}^{k}$ with the proof directly from
\cite[Theorem 9.10]{beck2014introduction}. 

In conclusion, once we obtain $\mathbf{w}^{k+1}=\mathbf{w}^{k}$ from
the proposed RFPA algorithm, $\mathbf{w}^{k}$ is a stationary point
of Problem (\ref{eq:MVSK}).
\end{IEEEproof}
\bibliographystyle{IEEEtran}
\bibliography{highOrderPortfolio}
 
\end{document}